\documentclass[aps,prb,reprint,superscriptaddress,
amsmath,citeautoscript,flushbottom,floatfix]{revtex4-1}

\usepackage{graphicx}

\newcommand{\vc}[1]{\boldsymbol{#1}}

\setcitestyle{numbers,square}

\allowdisplaybreaks

\begin{document}
\title{Highly frustrated magnetism in relativistic 
$\boldsymbol{d}^\mathbf{4}$ Mott insulators:\\ 
Bosonic analog of Kitaev honeycomb model}

\author{Ji\v{r}\'{\i} Chaloupka}
\affiliation{Department of Condensed Matter Physics, Faculty of Science,
Masaryk University, Kotl\'a\v{r}sk\'a 2, 61137 Brno, Czech Republic}
\affiliation{Central European Institute of Technology,
Masaryk University, Kamenice 753/5, 62500 Brno, Czech Republic}

\author{Giniyat Khaliullin}
\affiliation{Max Planck Institute for Solid State Research,
Heisenbergstrasse 1, D-70569 Stuttgart, Germany}

\begin{abstract}
We study the orbitally frustrated singlet-triplet models that emerge in the
context of spin-orbit coupled Mott insulators with $t_{2g}^4$ electronic
configuration. In these compounds, low-energy magnetic degrees of freedom can
be cast in terms of three-flavor ``triplon'' operators describing the
transitions between spin-orbit entangled $J=0$ ionic ground state and excited
$J=1$ levels. In contrast to a conventional, flavor-isotropic O(3)
singlet-triplet models, spin-orbit entangled triplon interactions are
flavor-and-bond selective and thus highly frustrated. In a honeycomb lattice,
we find close analogies with the Kitaev spin model -- an infinite number of
conserved quantities, no magnetic condensation, and spin correlations being
strictly short-ranged. However, due to the bosonic nature of triplons, there
are no emergent gapless excitations within the spin gap, and the ground state 
is a strongly correlated paramagnet of dense triplon pairs with no long-range
entanglement. Using exact diagonalization, we study the bosonic Kitaev model
and its various extensions, which break exact symmetries of the model and
allow magnetic condensation of triplons. Possible implications for magnetism
of ruthenium oxides are discussed.
\end{abstract}

\date{\today}

\maketitle


\section{Introduction}

Frustrated magnets where competing exchange interactions result in exotic
orderings and spin-liquid phases~\cite{Bal10,Sav17,Voj18} has been a subject
of active research over the years. In general, the magnetic moments in solids
are composed of spin and orbital angular momentum, with rather different
symmetry properties of interactions in spin and orbital sectors. While the
spin-exchange processes are described by isotropic Heisenberg model, the
orbital moment interactions are far more complex -- they are strongly
anisotropic in real and magnetic spaces \cite{Kug82,Kha05,Nus15} and
frustrated even on simple cubic lattices. The physical origin of this
frustration is that the orbitals are spatially anisotropic and hence cannot
simultaneously satisfy all the interacting bond directions in a crystal.

In late transition metal ion compounds, the bond-directionality and
frustration of the orbital interactions are inherited by the total angular
momentum $J=L+S$ \cite{Kha05}. Consequently, the low-energy ``pseudospin''
$J$-models may obtain nontrivial symmetries and host exotic ground states. The
best example of this sort is the emergence of the Kitaev honeycomb model
\cite{Kit06} in spin-orbit coupled Mott insulators of transition metal ions
with low-spin \mbox{$d^5$($S$=1/2, $L$=1)} \cite{Jac09,Cha10} and high-spin
\mbox{$d^7$($S$=3/2, $L$=1)} \cite{Liu18,San18} electronic configurations,
both possessing pseudospin $J=1/2$ Kramers doublet in the ground state.

The present paper studies the consequences of orbital frustration in another
class of spin-orbit Mott insulators, which are based on transition metal ions
with low-spin \mbox{$d^4$($S$=1, $L$=1)} electronic configuration such as
$4d$-Ru$^{4+}$ and $5d$-Ir$^{5+}$. In these compounds, spin-orbit coupling
$\lambda \vc L \cdot \vc S$ favors non-magnetic $J=0$ ionic ground state, and
magnetic order -- if any -- is realized via the condensation of excited $J=1$
triplet states \cite{Kha13,Mee15}. Near a magnetic quantum critical point,
where spin-orbit coupling and exchange interactions are of a similar strength
and compete, magnetic condensate can strongly fluctuate both in phase and
amplitude, as it has been observed in $d^4$ Mott insulator
\mbox{Ca$_2$RuO$_4$} \cite{Jai17,Sou17}.  

A minimal low-energy model describing the $J=0$ Mott insulators is a
singlet-triplet model, which can be written down in terms of three-flavor
``triplon'' operators $T_\alpha$ with $\alpha=x,y,z$ \cite{Kha13}. Distinct
from a conventional triplet excitations in spin-only models, the spin-orbit
entangled triplons keep track of the spatial shape of the $t_{2g}$ orbitals.
Therefore, their dynamics is expected to be flavor-and-bond selective and
frustrate the triplon condensation process. In broader terms, $J=0$ Mott
insulators provide a natural route to a phenomenon of frustrated magnetic
criticality \cite{Voj18}. 

The bond-directional nature of triplon dynamics is most pronounced in
compounds with $90^{\circ}$-exchange geometry as, e.g., in honeycomb lattice
\mbox{Li$_2$RuO$_3$} \cite{Miu07} with RuO$_6$ octahedra sharing the edges.
Previous work \cite{Kha13,Ani19} has already addressed singlet-triplet
honeycomb models, and found that the frustration effects can strongly delay
triplon condensation, or suppress it completely in the limit when only one
particular triplon flavor out-of-three $T_\alpha$ is active on a given bond.
Here, we perform a comprehensive symmetry analysis and exact diagonalization
of the model in this limit, where it features a number of properties of Kitaev
model.  For instance, we observe that the model has an extensive number of
conserved quantities, magnetic correlations are highly anisotropic and
confined to nearest-neighbor sites. We also find that the model is closely
related to the bilayer spin-1/2 Kitaev model \cite{Tom18,Sei18,Tom19}.
However, unlike the Kitaev spin-liquid with emergent non-local excitations,
the ground state of the model is a strongly correlated paramagnet smoothly
connected to the non-interacting triplon gas, and the lowest excitations are
of a single-triplon character at any strength of the exchange interactions. 

We analyze the model behavior also in the parameter regime where
singlet-triplet level is reversed (formally corresponding to the sign-change
of spin-orbit coupling), and find that triplon pairs condense into a
valence-bond-solid (VBS). This state is identical to the plaquette-VBS phase
of hard-core bosons on kagom\'e lattice \cite{Zha18}; this is not accidental,
since the symmetry properties of the model allow a mapping of triplon-pair
configurations on honeycomb lattice to a system of spinless bosons on dual
kagom\'e lattice. Further, adding the terms that relax the model symmetries,
we find a rich phase diagram including the magnetic and quadrupolar orderings. 

The paper is organized as follows: Sec.~\ref{sec:model} introduces the model
and sketches its derivation. In Sec.~\ref{sec:symm} we analyze the model
symmetries and find analogies to the Kitaev honeycomb model. The phase
diagrams and spin excitations are studied in Secs.~\ref{sec:chain} and
\ref{sec:honey} -- for the simpler one-dimensional (1D) analog of the model
providing useful insights, and the full model on the honeycomb lattice,
respectively. Sec.~\ref{sec:concl} summarizes the main results. 


\section{Honeycomb singlet-triplet model}
\label{sec:model}

\begin{figure}[tb]
\begin{center}
\includegraphics[scale=0.97]{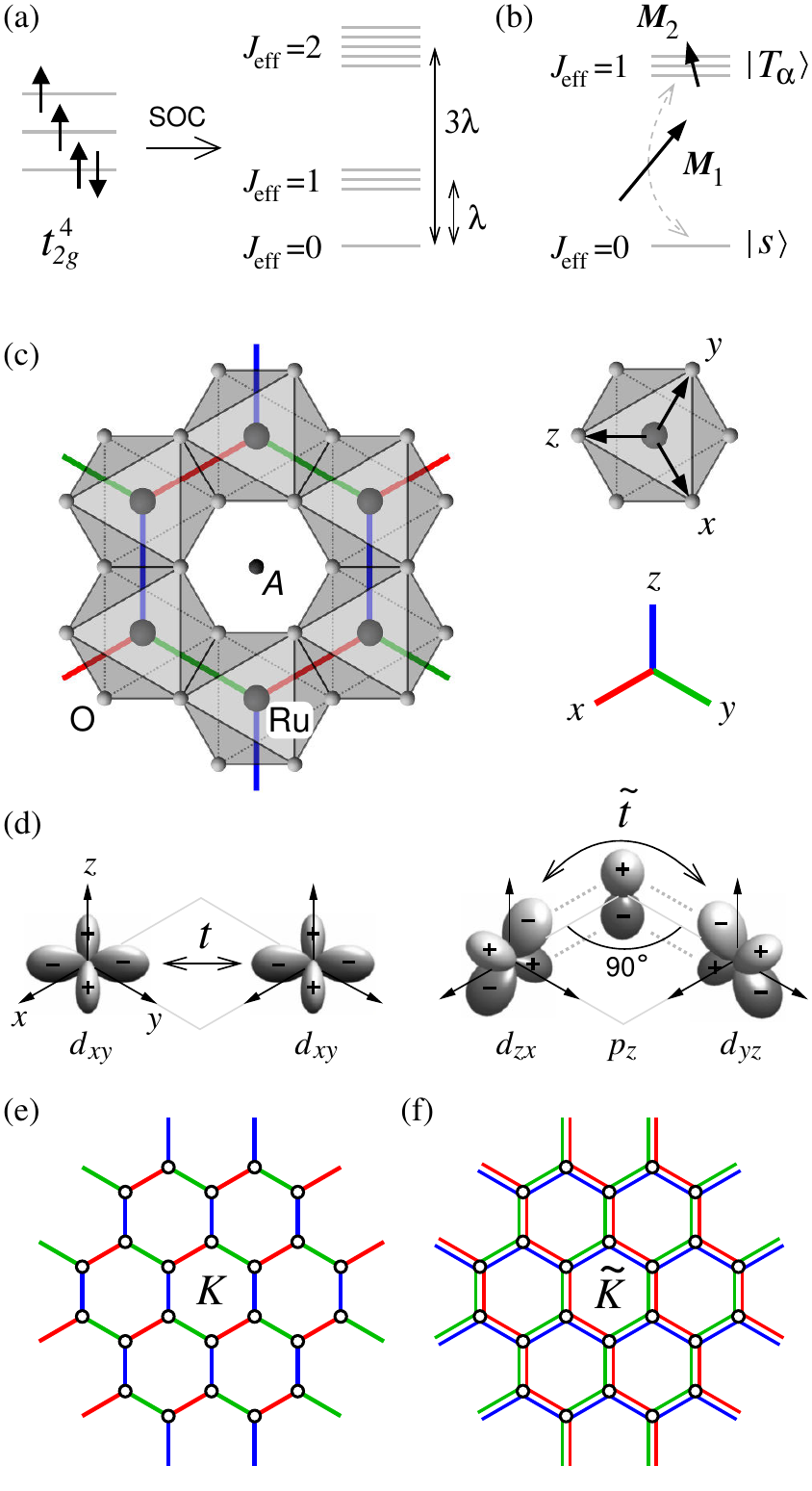}
\caption{
(a)~Energy levels of $t_{2g}^4$ configuration in $LS$ coupling scheme.
Low-energy $J_\mathrm{eff}=0$ singlet and $J_\mathrm{eff}=1$ triplet form the
local basis of our effective model.
(b)~The magnetic moment consists of dominant van Vleck-type contribution $\vc
M_1$ residing on the $J_\mathrm{eff}=0\leftrightarrow 1$ transition, and a
smaller contribution $\vc M_2$ carried by $J_\mathrm{eff}=1$ triplet.
(c)~Top view of the honeycomb lattice of edge-shared RuO$_6$ octahedra in
$A_2$RuO$_3$. Cubic axes $x$, $y$, $z$ pointing from Ru towards O ions, as
well as the three types (red, green, blue) of nearest-neighbor bonds
in the honeycomb lattice are indicated.
(d)~Hopping within Ru$_2$O$_2$ plaquette of a $z$-bond proceeds through direct
overlap of $d$ orbitals (left) or indirectly via oxygen ions (right).
(e)~Kitaev-like pattern of active bond colors for the interaction $K$ in
Eq.~(\ref{eq:Ham}) in the direct-hopping case. On $z$-bonds, the blue-color
triplons $T_z$ are active, etc.  
(f)~Complementary $xy$-type pattern for the interaction $\tilde K$ emerging in
the case of indirect hopping; on $z$-bonds, the red ($T_x$) and green ($T_y$)
color triplons are active.
}\label{fig:schem}
\end{center}
\end{figure}

We consider a transition metal ion with four electrons on $t_{2g}$ level, e.g.
Ru$^{4+}$. Spin-orbit coupling results in a multiplet structure shown in
Fig.~\ref{fig:schem}(a). A minimal model for magnetism of such van Vleck-type
ions includes the lowest excited $J_\mathrm{eff}=1$ states $|T_{\pm 1}\rangle$
and $|T_0\rangle$, in addition to the ground state $J_\mathrm{eff}=0$ singlet
$|s\rangle$. It is convenient to work with three triplon operators $T_\alpha$
of Cartesian flavors (``colors'') $\alpha=x,y,z$. Using the above $J_z$
eigenstates, they are defined as
\begin{equation}
T_x = \tfrac1{i\sqrt2}( T_1-T_{-1} ) , \;
T_y = \tfrac1{\sqrt2} ( T_1+T_{-1} ) , \; 
T_z = i T_0,
\end{equation}
and together form a Cartesian vector $\vc T$. A constraint $n_x+n_y+n_z+n_s=1$ 
with $n_\alpha=T^\dagger_\alpha T_\alpha$ and $n_s=s^\dagger s$ is implied.
Spin-orbit splitting reads then as a chemical potential for $T_\alpha$ bosons:
$\lambda\, n_T = \lambda\,(n_x+n_y+n_z)$.

As illustrated in Fig.~\ref{fig:schem}(b), local magnetic moment is composed
of two terms, $\vc M=\vc M_1+\vc M_2$, where $\vc M_1$ originates from
dipolar-active transitions between $s$ and $T_\alpha$ states \cite{Kha13}: 
\begin{equation}\label{eq:M1}
\vc M_1 = 2\sqrt{6}\, \vc v = -\sqrt6\, i \,
(s^\dagger \vc T -\vc T^\dagger s), 
\end{equation}
while $\vc M_2$ is derived from triplon-spin $\vc J=-i(\vc T^\dagger\times\vc
T)$ with $g$-factor $1/2$:  
\begin{equation}\label{eq:M2}
\vc M_2 = \tfrac12\vc J = -\tfrac 12\, i\, 
(\vc T^\dagger \times \vc T). 
\end{equation}
In Eq.~\eqref{eq:M1}, $\vc v=-\frac12i(s^\dagger \vc T -\vc T^\dagger s)$
keeps track of the imaginary part of $\vc T$ (real part of $\vc T$ carries a
quadrupolar moment).

Triplon interactions are derived from Kugel-Khomskii-type exchange
Hamiltonian, projected onto singlet-triplet basis \cite{Kha13,Ani19}. In
honeycomb lattice of the edge-shared RuO$_6$ octahedra, see
Fig.~\ref{fig:schem}(c), there are two types of electron exchange processes,
generated by (i) a direct hopping $t$ of $d$-electrons between
nearest-neighbor (NN) Ru ions, and (ii) indirect hopping $\tilde t$ via oxygen
ions, as depicted in Fig.~\ref{fig:schem}(d). Consider, for example, direct
$t$-hopping; for a $z$-type bond, it reads as 
$-t(d_{xy,i}^\dagger d_{xy,j}^{\phantom{\dagger}}+\mathrm{H.c.})$. 
Second order perturbation theory gives the exchange Hamiltonian, written in
terms of spin $S=1$ and orbital $L=1$ operators of $d^4$ configuration:
\begin{equation}\label{eq:KK}
\mathcal{H}_{ij}^{(z)} = \frac{t^2}{U}\,\left[(\vc S_i\cdot\vc S_j+1) 
L_{zi}^2L_{zj}^2\!-\!L_{zi}^2\!-\!L_{zj}^2\right]. 
\end{equation}
Next, one has to calculate the matrix elements of operators in
Eq.~\eqref{eq:KK} between $J_\mathrm{eff}=0$ and $J_\mathrm{eff}=1$
wavefunctions \cite{Kha13}, and represent them in terms $T_\alpha$ and $s$.
For example, $S^zL_z^2=\sqrt{\frac{8}{3}}\; v_z$, while
$S^{x/y}L_z^2=\sqrt{\frac{2}{3}}\; v_{x/y} + \frac{1}{2}J_{x/y}$, with
``van-Vleck'' moments $\vc v$ and triplon-spin $\vc J$ already defined above.
The projected Hamiltonian \eqref{eq:KK} takes the form of  
$\mathcal{H}_{ij}^{(z)}=\frac83 \frac{t^2}{U}(h_2+h_3+h_4)^{(z)}_{ij}$. 
It contains two-, three-, and four-triplon operator terms: 
\begin{align}
h_2^{(z)} &=v_{zi}v_{zj} +
            \frac14 (v_{xi}v_{xj}+v_{yi}v_{yj}) , \label{eq:h2} \\
h_3^{(z)} &=\frac18 \sqrt{\frac32} (v_{xi}J_{xj}+v_{yi}J_{yj}) 
            + (i\leftrightarrow j) , \label{eq:h3}\\
h_4^{(z)} &=\frac3{32} (J_{xi}J_{xj}+J_{yi}J_{yj}) 
            + \frac1{32}\, Q_{3i}Q_{3j} , \label{eq:h4}
\end{align}
where $Q_3=(n_x+n_y-2n_z)/\sqrt{3}$ is a quadrupole operator of $E_g$
symmetry. Interactions $\mathcal{H}_{ij}^{(x)}$ and $\mathcal{H}_{ij}^{(y)}$
for $x$- and $y$-type bonds follow from symmetry. The largest term in the
above Hamiltonian is represented by $v_{zi}v_{zj}$ coupling in $h_2$;
physically, this is Ising-type coupling between van-Vleck moments.

Indirect hopping via ligands 
$-\tilde{t}(d_{yz,i}^\dagger d_{zx,j}^{\phantom{\dagger}}+\mathrm{H.c.})$
generates triplon Hamiltonian of the same form, 
$\mathcal{\tilde H}_{ij}^{(z)}=3\frac{\tilde t^2}{U}
(\tilde h_2+\tilde h_3+\tilde h_4)^{(z)}_{ij}$. In contrast to the above case, 
however, a dominant term here is represented by $xy$-type coupling 
$(v_{xi}v_{xj}+v_{yi}v_{yj})$ in $\tilde h_2$ (explicit forms of the other 
terms can be found in Ref.~\cite{Kha13}).  

The full models $\mathcal{H}$ and $\mathcal{\tilde H}$ are clearly rich but
complicated; considering their dominant terms represented by Ising- and
$xy$-type couplings between $v$-moments should provide some useful insights.
Even though these couplings look as simple quadratic forms, the hard-core
nature of triplons and their bond-directional anisotropy lead to nontrivial
consequences \cite{Kha13,Ani19}. 

We introduce the bond operator $\mathcal{O}_{ij}^{\alpha}=4v_{\alpha i}
v_{\alpha j}$, which in terms of singlet $s$ and triplon $T_\alpha$ operators
reads as: 
\begin{equation}\label{eq:Hcolor}
\mathcal{O}_{ij}^{\alpha} = 
T^\dagger_{\alpha i} s^{\phantom{\dagger}}_i 
s^\dagger_j T^{\phantom{\dagger}}_{\alpha j} - 
T^\dagger_{\alpha i} s^{\phantom{\dagger}}_i 
T^\dagger_{\alpha j} s^{\phantom{\dagger}}_j + \mathrm{H.c.}
\end{equation}
We recall that $s$ and $T_\alpha$ are subject to local constraint $n_s+n_T=1$.
Alternatively,
\begin{equation}\label{eq:Hcolor}
\mathcal{O}_{ij}^{\alpha} = 
\mathcal{T}^\dagger_{\alpha i} 
\mathcal{T}^{\phantom{\dagger}}_{\alpha j} - 
\mathcal{T}^\dagger_{\alpha i} 
\mathcal{T}^\dagger_{\alpha j} + \mathrm{H.c.}, 
\end{equation}
where $\mathcal{T}^\dagger= T^\dagger s$ is a hard-core boson with 
$n_{\mathcal T}\leq 1$.
In terms of $\mathcal{O}_{ij}^{\alpha}$, a minimal singlet-triplet model
Hamiltonian can concisely be written as 
\begin{equation}\label{eq:Ham}
\mathcal{H} = \sum_i E_T\, n_{T_i} +
\sum_{{\langle ij\rangle}_\alpha}
[K\mathcal{O}_{ij}^{\alpha} 
+\tilde{K}(\mathcal{O}_{ij}^{\bar{\alpha}}
          +\mathcal{O}_{ij}^{\bar{\bar{\alpha}} }) ].  
\end{equation}
Here, the color $\alpha\in\{x,y,z\}$ is given by the direction of the 
bond $\langle ij\rangle$, and $\bar{\alpha}$, $\bar{\bar{\alpha}}$ are the
two complementary colors; e.g., for $z$-type bond $\langle ij\rangle_z$ one
has $\alpha=z$, while $\bar{\alpha}=x$ and $\bar{\bar{\alpha}}=y$.  
As derived, the model parameters are $E_T=\lambda$, $K=\frac23 t^2/U$, 
$\tilde{K}=\frac14 K+\frac34 \tilde{t}^{\:2}/U$. 

The $K$ ($\tilde K$) term in Eq.~\eqref{eq:Ham} features Kitaev-like
(\mbox{$xy$-type}) symmetry, with one (two) active components of \mbox{$\vc
T$-vector} on a given bond. The resulting color-and-bond selective interaction
patterns $K$ and $\tilde K$ are shown in Fig.~\ref{fig:schem}(e) and
Fig.~\ref{fig:schem}(f), correspondingly. At $K=\tilde{K}$, the model is
equivalent to a conventional O(3) singlet-triplet models \cite{Gia08} that
appear, e.g., in low-energy description of a bilayer Heisenberg system. In
this isotropic limit, the model is free of frustration and would undergo a
magnetic transition at large enough coupling strength $K \sim E_T$. In this
paper, the Kitaev-like model with $K$-term, where triplon dynamics is most
frustrated, is of primary interest. In particular, we are interested in the
nature of magnetically disordered ground state at strong coupling limit of
$K\gg E_T$.  In real materials, an admixture of the complementary interaction
$\tilde{K}$ is expected, and we will consider its impact on the phase behavior
of the model. 


\section{Symmetry properties and links to Kitaev honeycomb model}
\label{sec:symm}

The color--bond correspondence of the above model in the $K\neq 0$,
$\tilde{K}=0$ case is strongly reminiscent of the Kitaev honeycomb model. In
this section, we focus exclusively on this limit, draw the corresponding
analogies, and find an exact link between our model and a particular variant
of bilayer Kitaev model.

\subsection{Extensive number of conserved quantities}
\label{sec:locZ2}

In the Kitaev-like limit of the model in Eq.~\eqref{eq:Ham}, i.e.
$\tilde{K}=0$, the number of $T_\alpha$ stays either even ($0$ and $2$) or odd
($1$) on a bond of direction $\alpha$. The parity of this number is thus a
conserved $Z_2$ quantity that can be formally written as
\begin{equation}
P_{ij} = (1-2n_{\alpha i})(1-2n_{\alpha j})
\end{equation}
with $n_{\alpha i}$ counting $T_\alpha$-triplon number on site $i$. Being
associated with the individual bonds, the parities form an extensive set of
conserved $Z_2$ quantities that decompose the Hilbert space into subspaces
with fixed bond-parity configurations. The total Hilbert space dimension
equals $4^N$ for a system with $N$ sites. With one $Z_2$ conserved quantity
per bond (amounting to $\frac 32$ per site), the average subspace dimension is
reduced to $4^N/2^\frac{3N}2 = (\sqrt{2})^N$. This is actually the same
scaling as in the case of the Kitaev honeycomb model \cite{Kit06}, where the
conserved $Z_2$ quantities are associated with hexagonal plaquettes (giving
$\frac12$ of $Z_2$ quantity per site) and the average subspace dimension thus
becomes $2^N/2^\frac{N}2 = (\sqrt{2})^N$. In accord with intuitive
expectation, our numerical calculations found the ground state to have {\em
all-even} bond-parity configuration.

\subsection{Mapping to hardcore bosons on a dual lattice}
\label{sec:dualmap}

When working in the subspaces with fixed bond parities, most of the $4^N$
configurations of triplons on the honeycomb lattice of size $N$ are
irrelevant. To remove this redundancy in the description, here we develop an
auxiliary particle representation by mapping to a system of spinless hardcore
bosons on dual (kagom\'e) lattice.

\begin{figure}[b]
\begin{center}
\includegraphics[scale=1.00]{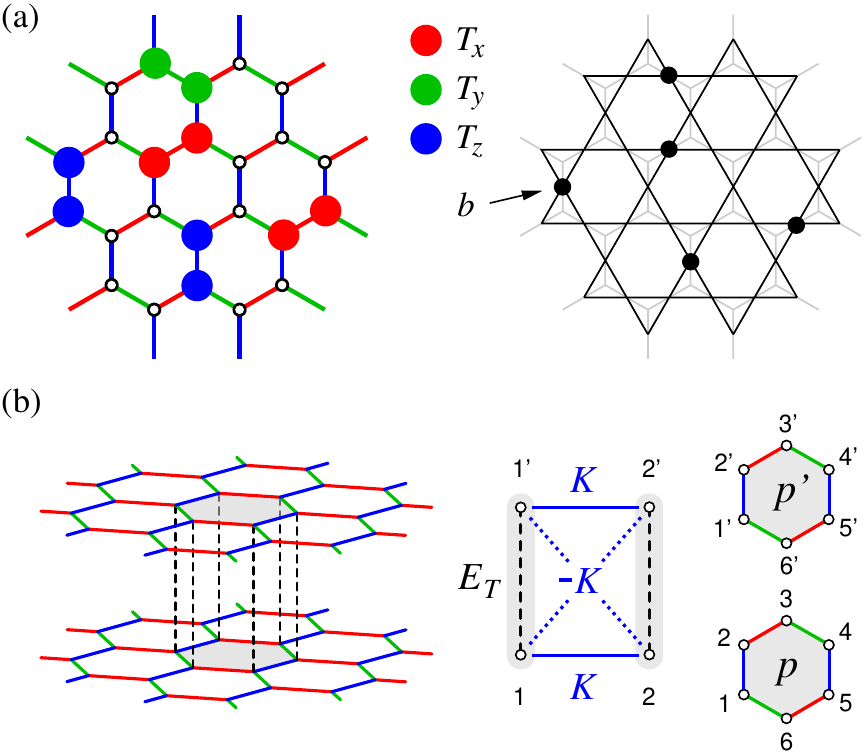}
\caption{
(a)~Sample triplon configuration on the honeycomb lattice (left) and the
corresponding configuration of spinless hardcore bosons $b$ (black dots)
on a dual kagom\'e lattice (right). In the case of all-even bond parities 
depicted here, the presence of a $T$-dimer with a proper color (i.e. same 
as that of the bond) is represented by $b$ boson on the dual lattice.
No nearest-neighbor pairs of $b$ on a kagom\'e lattice is allowed. 
(b)~Equivalent \mbox{spin-$\frac12$} bilayer model realizing the
singlet-triplet basis on interlayer vertical bonds that are subject 
to Heisenberg interaction $J=E_T$. 
In the model, Kitaev interaction $K$ active on intralayer bonds (such as $1-2$
and $1'-2'$) is complemented by a Kitaev interaction $-K$ of opposite sign,
acting on interlayer cross links (such as $1-2'$ and $1'-2$).
Vertically adjacent hexagonal plaquettes $p$, $p'$ are used to construct 
the conserved quantities.
}\label{fig:mappings}
\end{center}
\end{figure}

For simplicity, we limit ourselves to the case of {\em all-even} bond
parities, similar one-to-one mappings can be found also in the other cases.
The mapping is illustrated in Fig.~\ref{fig:mappings}(a). A given bond of the
honeycomb lattice can either be occupied by a pair of triplons of the proper
color or be empty. These two states are represented by the presence/absence of
a hardcore boson $b$ on the corresponding dual lattice site. Starting from a
$b$-configuration, the state of a given honeycomb site can be uniquely
reconstructed by checking the surrounding kagom\'e sites for $b$ bosons.
Either (i) one of them is found, selecting one of the $T_\alpha$ states with
$\alpha$ depending on the bond occupied by the $b$ boson, or (ii) none is
present, corresponding to the ``empty'' states $s$. The constraint for the $b$
bosons is now evident -- a nearest-neighbor pair of $b$ at the dual lattice is
forbidden. 

Altogether, we can formulate the Hamiltonian for the $b$ bosons on the dual
lattice as
\begin{equation}\label{eq:Hdual}
\mathcal{H}_\mathrm{dual} = \sum_i [2E_T n_b - K (b + b^\dagger)]_i
+ U\sum_{\langle ij\rangle} n_{bi} n_{bj} ,
\end{equation}
where $i$ runs through the sites of the kagom\'e lattice, and the repulsive
interaction with $U\rightarrow \infty$ enforces the constraint of ``no
nearest-neighbor occupation'' for $b$ bosons. Without this constraint, the sum
of local Hamiltonians in \eqref{eq:Hdual} would be easy to diagonalize leading
to bond eigenstates that involve $|ss\rangle$ and $|T_\alpha T_\alpha\rangle$
pairs. In the form of Eq.~\eqref{eq:Hdual}, the peculiarity of the model is
fully exposed -- the $K$ interaction forms bond dimers that communicate via
constraint only. Adding intersite hopping terms $b^\dagger_ib_j$ in the model
would generate boson dispersion and phase relations between them on different
sites, leading to a superfluid condensate; however, we have so far no clear
microscopic mechanism that would result in such a triplon-pair hopping
process. 

Finally, let us recall that the above $\mathcal{H}_\mathrm{dual}$ is valid for
the all-even sector only; the formulation of the constraint in the other cases
is more complicated. 

\subsection{Local nature of the spin correlations}
\label{sec:spincorr}

Similarly to the Kitaev model, the presence of the local conserved quantities
has consequences for the spin correlations, both of van Vleck moments of
Eq.~\eqref{eq:M1} as well as triplon spins entering Eq.~\eqref{eq:M2}.
Let us consider static correlations of the type
$\langle M_{\alpha} M_{\alpha} \rangle = Z^{-1} \sum_n
\langle n | M_{\alpha} M_{\alpha} |n\rangle\, \mathrm{e}^{-\beta E_n}$
or the corresponding dynamic correlations. The eigenstates $|n\rangle$ of
Eq.~\eqref{eq:Ham} with $\tilde{K}=0$ have fixed bond-parity configurations.
When acting by the $\alpha$-component of the van Vleck moment operator
$M_{1\alpha} \propto (s^\dagger T^{\phantom{\dagger}}_\alpha-T^\dagger_\alpha s)$ 
on a given site, the bond parity of the attached $\alpha$-bond is switched. 
Bond parities are conserved by the Hamiltonian, the introduced parity defect
is thus immobile, and to get back to original parity configuration, one has to
act with $M_{1\alpha}$ either at the same site or on the second site of the
affected $\alpha$-bond. Therefore, $\langle M_{\alpha} M_{\alpha} \rangle$
correlator is strictly zero beyond a nearest-neighbor distance. Similarly,
the triplon spin operator $-i(\vc T^\dagger \times \vc T)$ flips parities of
two attached bonds, the original bond-parity configuration therefore has to be
restored by acting at the same site. As a result, the Kitaev-like limit
of the model is characterized by nearest-neighbor only correlations of the
magnetic moments (stemming from the van Vleck component matching the bond
color), and a localized nature of the dynamic spin response. This mechanism
is completely analogous to the Kitaev model, where a spin flip introduces
two localized fluxes \cite{Kit06}. 

\subsection{Links to the Kitaev honeycomb model}
 
In the previous Secs. \ref{sec:locZ2} and \ref{sec:spincorr} we have
noticed several striking similarities between the bosonic $K$-model and
Kitaev's model for \mbox{spins-$\frac12$} residing on the honeycomb lattice. 
A deeper connection of the two models can be thus anticipated, motivating the
search for a \mbox{spin-$\frac12$} equivalent of our model that could reveal
such a  link. A natural search direction is the class of bilayer
\mbox{spin-$\frac12$} systems with Heisenberg interlayer interaction forming
a local singlet-triplet basis on the interlayer rungs.

Indeed, the Hamiltonian in Eq.~\eqref{eq:Ham} can be exactly mapped onto
\mbox{spin-$\frac12$} bilayer honeycomb system with the interactions $K$ and
$\tilde{K}$ transforming into nearest-neighbor intralayer links and second
nearest-neighbor interlayer links as depicted in Fig.~\ref{fig:mappings}(b).
For $\tilde{K}=0$, the Hamiltonian involving the nearest-neighbor bond
\mbox{$1$-$2$} and the adjacent one \mbox{$1'$-$2'$} in the other layer reads
as
\begin{multline}\label{eq:Hbil}
\mathcal{H}_{121'2'} = 
E_T (\vc S_1 \vc S_{1'}+\vc S_2 \vc S_{2'}) \\
+K (S^\alpha_1 S^\alpha_2+S^\alpha_{1'} S^\alpha_{2'})
-K (S^\alpha_1 S^\alpha_{2'}+S^\alpha_{1'} S^\alpha_2) .
\end{multline}
The first two terms form nothing but a pair of Kitaev models linked by
vertical Heisenberg bonds. This so-called bilayer Kitaev model was recently
studied in Refs.~\cite{Tom18,Tom19,Sei18}. The last term in Eq.~\eqref{eq:Hbil}
provides additional Kitaev-like cross-links of the sign opposite to the
intra-layer Kitaev interaction and, as we find later, drastically changes
the behavior of the system from that of standard bilayer Kitaev model.

With the above mapping, we are ready to consider the relations between various 
local conserved quantities. The single layer Kitaev model conserves the product
of spin operators at a hexagonal plaquette [see Fig.~\ref{fig:mappings}(b)]
\begin{equation}
W_p = 2^6 S^x_1 S^y_2 S^z_3 S^x_4 S^y_5 S^z_6 .
\end{equation}
In a bilayer Kitaev model, one has to construct products $W_p W_{p'}$ of
Kitaev's $W_p$ for vertically adjacent plaquettes \cite{Tom18}. These
conserved $Z_2$ quantities bring about certain features of Kitaev physics to
the bilayer Kitaev model. In the case of our model~\eqref{eq:Hbil}, the extra
Kitaev-like cross-links are present. However, the products $W_p W_{p'}$ are
still conserved, as can be verified by a direct calculation. Surprisingly,
this does not make them yet another set of conserved quantities. In fact, it
turns out that $W_p W_{p'}$ are merely products of our bond parities $P_{ij}$
\begin{equation}
W_p W_{p'} = P_{12} P_{23} P_{34} P_{45} P_{56} P_{61}
\end{equation}
in the original formulation, and appear as a simple consequence of the
bond-parity conservation in the model. The above connection also translates
the all-even parity configuration of the ground state into the absence of
fluxes in the ground state, i.e. $W_p W_{p'}=+1$ for all plaquettes. The local
symmetries of our model are thus more powerful than in the Kitaev model or its
simple bilayer extension.  Intuitively it may be expected that this denser
covering by local conserved quantities will lead to less entangled (more
factorized) ground states, as we indeed find below. 

As a side remark, we note that while the Hamiltonian \eqref{eq:Ham} contains a
balanced combination of hopping and pair terms, differing only by the sign
[see Eq.~\eqref{eq:Hcolor}], it is possible to generalize the above mapping to
the case  $A\, T^\dagger_\alpha T_\alpha - B\, T^\dagger_\alpha
T^\dagger_\alpha + \mathrm{H.c.}$ with $A\neq B$. The resulting
\mbox{spin-$\frac12$} interactions consist of $\mathcal{H}_{121'2'}$ of
Eq.~\eqref{eq:Hbil} with $K=\tfrac12(A+B)$ and an additional four-spin
interaction
\begin{equation}\label{eq:Hbilgen}
\Delta\mathcal{H}_{121'2'}=
2(A-B)
(\vc S_1\times\vc S_{1'})^\alpha
(\vc S_2\times\vc S_{2'})^\alpha .
\end{equation}
By introducing symmetric off-diagonal exchange $\Gamma_{ij}=S_i^x S_j^y +
S_i^y S_j^x$ (for a $z$ bond, $x$ and $y$-bond expressions are obtained by
cyclic permutation), it can be brought to a form
$2(B-A)(\Gamma_{12}\Gamma_{1'2'}-\Gamma_{12'}\Gamma_{1'2})$ resembling
somewhat the structure in Eq.~\eqref{eq:Hbil}.
All the arguments concerning conserved quantities remain valid also in the
$A\neq B$ case, because the original interactions in expressed using the $T$
particles manifestly conserve bond parities. For example, despite the
complicated structure of Eq.~\eqref{eq:Hbilgen}, it commutes with the plaquette
products $W_p W_{p'}$ keeping them still conserved.


\section{Kitaev-like singlet-triplet zigzag chain}
\label{sec:chain}


Before studying the full model on the honeycomb lattice, we first focus on its
one-dimensional analog. The 1D system is more accessible to numerics and
enables easier insights. As an example of such an approach in the context of
the Kitaev-Heisenberg model, Ref.~\cite{Agr18} studies the corresponding 1D
chains and subsequently makes an interpretation of the 2D honeycomb model
behavior in terms of coupled 1D chains.

To form a 1D model analogous to the honeycomb one, we remove $T_z$ triplon and
keep only one zigzag chain of the honeycomb lattice, consisting of $x$ and $y$
bonds. In the Kitaev-like limit $\tilde{K}=0$, these two changes are
equivalent as only $T_z$ is active on the $z$ bonds. Going away from the
Kitaev-like limit, as an alternative to the complementary $\tilde{K}$
interaction, it is more transparent here to add bond-independent $J_\mathrm{XY}$
interaction. Instead of the model in Eq.~\eqref{eq:Ham} we therefore deal with
\begin{equation}\label{eq:Ham1D}
\mathcal{H} = \sum_i E_T\, n_{T_i} + 
\sum_{{\langle ij \rangle}_\alpha} [K \mathcal{O}_{ij}^\alpha +
J_\mathrm{XY}(\mathcal{O}_{ij}^x+\mathcal{O}_{ij}^y)]
\end{equation}
formulated for a zigzag chain of alternating $x$ and $y$ bonds with the bond
direction determining again the color $\alpha$. Note that for
$J_\mathrm{XY}\neq 0$, a slight change to the original parametrization occurs:
$K=(K-\tilde{K})_\mathrm{orig}$, $J_\mathrm{XY}=\tilde{K}_\mathrm{orig}$.

Having now only three levels $s$, $T_x$, $T_y$ in the 1D model, it is possible
to convert it to a \mbox{spin-$1$} chain using the transformation
\begin{align}
\label{eq:sTspin1a} S^x &= -i(s^\dagger T^{\phantom{\dagger}}_x - T^\dagger_x s) , \\
\label{eq:sTspin1b} S^y &= -i(s^\dagger T^{\phantom{\dagger}}_y - T^\dagger_y s) , \\
\label{eq:sTspin1c} S^z &= -i(T^\dagger_x T^{\phantom{\dagger}}_y -
T^\dagger_y T^{\phantom{\dagger}}_x) ,
\end{align}
where the first two components of the effective \mbox{spin-$1$} correspond to
van Vleck moments $\vc v$ while the last one is linked to the triplon spin
$\vc J$ [see Eqs.~\eqref{eq:M1} and \eqref{eq:M2}]. The resulting equivalent
\mbox{spin-$1$} model is a Kitaev-XY \mbox{spin-$1$} chain with single-ion
anisotropy $E_T$:
\begin{equation}\label{eq:Ham1Dspin1}
\mathcal{H} = \sum_i E_T (S_i^z)^2 +
\sum_{{\langle ij \rangle}_\alpha}[ K S_i^\alpha S_j^\alpha + 
J_\mathrm{XY} (S_i^x S_j^x+S_i^y S_j^y)] .
\end{equation}
The phase diagram of this model for the $K=0$ case (no bond-alternation) was
thoroughly explored in the context of \mbox{spin-$1$} XXZ chain with single-ion
anisotropy (see, e.g., Refs.~\cite{Bot83,Sch86,Nij89,Che03}). Later in
Sec.~\ref{sec:chainKXY}, we will use these corresponding studies as a reference.


\subsection{Chain with pure Kitaev-like interaction}
\label{sec:chainK}

\begin{figure}
\begin{center}
\includegraphics[scale=1.00]{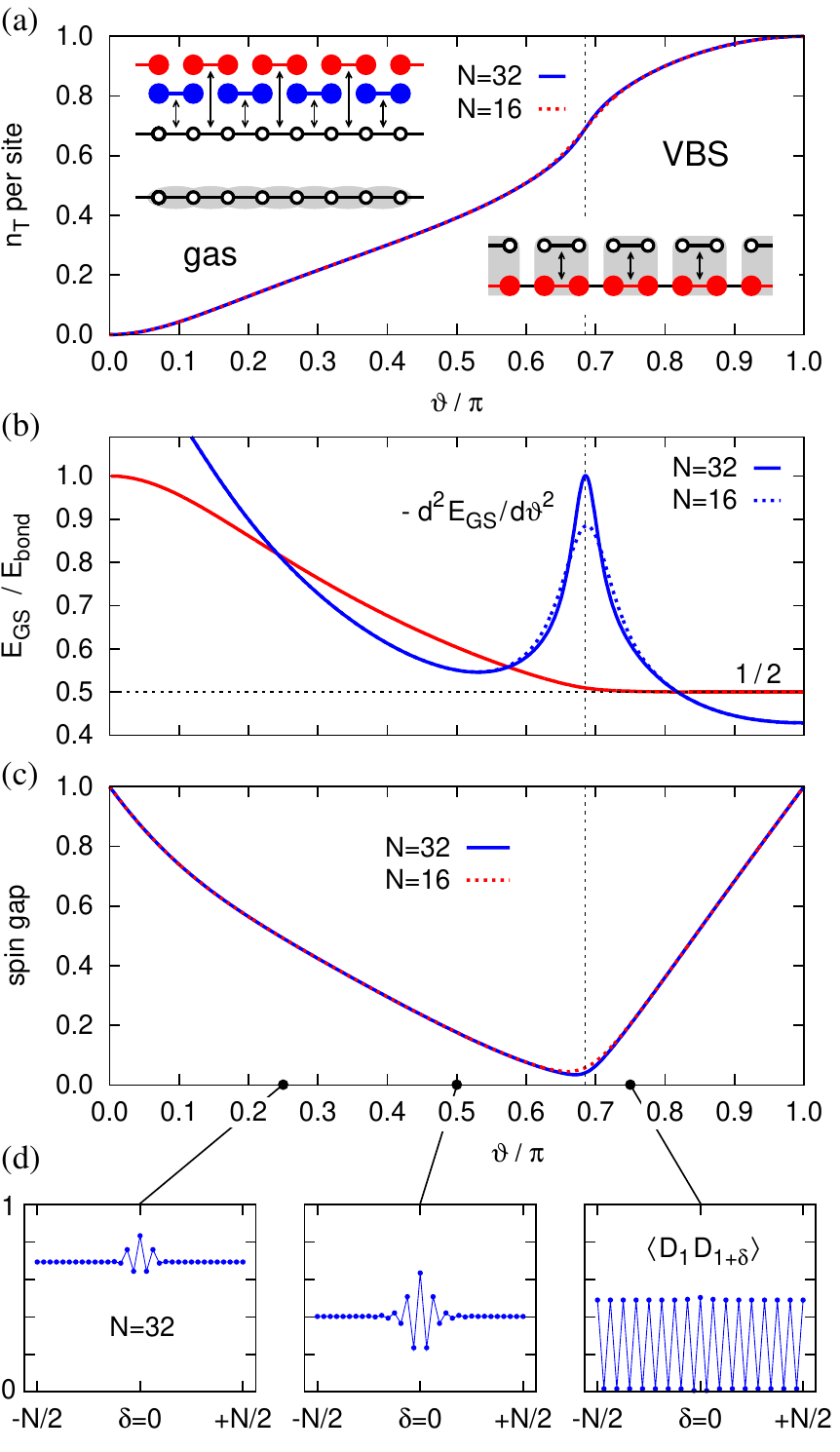}
\caption{
(a)~Occupation of the triplon states $T_{x,y}$ within the chain model
parametrized as $E_T=\cos\vartheta$, $K=\sin\vartheta$ obtained by exact
diagonalization.  Results for the chains of the length $N=16$ and $N=32$ are
nearly identical. The insets show a cartoon picture of the groundstate: At
positive $E_T\gg K$, each bond can be predominantly in the bonding state as it
is mostly composed of $s$. At negative $E_T$ with $|E_T|\gg K$, the bonding
states are incompatible and are realized only on every second bond creating
thus a valence bond solid.
(b)~Ground-state energy per bond $E_\mathrm{GS}$ measured by bonding state
energy $E_\mathrm{bond}$ (red) and the second derivative of $E_\mathrm{GS}$
with respect to $\vartheta$ revealing the quantum critical point (blue).
(c)~Spin gap obtained as the difference of ground-state energies within
all-even sector and the sector with a single odd-parity bond.
(d)~Examples of bond-dimer correlations $\langle D_1 D_{1+\delta}\rangle$ with
the dimer operator defined as a projector to the bonding state of
Eq.~\eqref{eq:bonding}: $D=|B\rangle\langle B|$.  The value of the correlator
at $\delta=0$ gives the probability of the bonding state $P_B$, at large
distances it approaches $P_B^2$ in the gas phase.  Oscillations near
$\delta=0$ growing with $\vartheta$ are due to the increasing incompatibility
of bonding states at adjacent bonds.  The correlation length diverges
approaching the QCP and long-range correlations are seen in the VBS phase.
}\label{fig:chainK}
\end{center}
\end{figure}

As the first step, we consider the Kitaev-like limit of the model
\eqref{eq:Ham1D}, i.e. neglecting $J_{XY}$ term. In general, the behavior of
all our models is determined by a competition of the triplon energy cost $E_T$
with the energy gain due to the interactions. One can thus expect a quantum
critical point (QCP) separating a triplon gas phase with small triplon
densities (dominant $E_T$ regime) from a phase characterized by strongly
interacting triplons at larger densities (dominant $K$ regime).

Such a competition is captured by Fig.~\ref{fig:chainK} presenting an
evolution for varying $K$ to $E_T$ ratio. For a better understanding and to
actually reach the QCP in this case, we have extended the parameter range to
the (non-realistic in the present physical context but interesting
theoretically) regime of $E_T<0$ with reversed $s$ and $T$ levels. The data
obtained by exact diagonalization (ED) are shown for two chain lengths to
assess the finite-size effects that are quite small here. As seen in
Fig.~\ref{fig:chainK}(a), for an increasing interaction strength $K$, the
triplon density $n_T$ gradually increases with just a single hint of a change
of the regime located already at negative $E_T$. This QCP is clearly revealed
by the second derivative of the ground-state energy $E_\mathrm{GS}$ with
respect to model parameters as presented in Fig.~\ref{fig:chainK}(b).

To understand the energetics of the evolution together with the nature of the
two phases, it is convenient to measure the chain $E_\mathrm{GS}$ per bond by
a ground-state energy of an isolated bond, as it is done in
Fig.~\ref{fig:chainK}(b). The ground state of a single bond -- the bonding
state -- mixes a pair of proper-color triplons and a pair of $s$ in the
wavefunction 
\begin{equation}\label{eq:bonding}
|B\rangle=\cos\phi\, |ss\rangle + \sin\phi\, |T_\alpha T_\alpha\rangle
\end{equation}
with $\phi$ given by $\tan2\phi=K/E_T$, and has the energy 
$E_\mathrm{bond} = E_T - \sqrt{E_T^2+K^2}$.
This approximately evaluates to $-K^2/2E_T$ for small $K$, capturing the
perturbative incorporation of a triplon pair (energy $2E_T$) by a process with
an amplitude $K$. The orthogonal combination to $|B\rangle$ is the antibonding
state 
$|A\rangle=\cos\phi\, |T_\alpha T_\alpha\rangle - \sin\phi\, |ss\rangle$ 
whose energy starts at $2E_T$ in $K\rightarrow 0$ limit.
Similarly to $n_T$, the ground-state energy measured by $E_\mathrm{bond}$
shows a gradual evolution with the model parameters for most of the parameter
range apart from a change at QCP [see Fig.~\ref{fig:chainK}(b)]. In the
$E_T\gg K$ regime, $E_\mathrm{GS}$ reveals the dominance of the bonding
states that seem to fill up the system. This is possible since the bonding
states are composed mostly of $s$ states that can be shared by the neighboring
bonds. For increasing $K$ and thus an increasing admixture of $T$ pairs with
bond-dependent color, the bonding states at neighboring bonds have less
overlap and the energy gain from $K$ is reduced compared to that of isolated
bonds. At the QCP, $E_\mathrm{GS}/E_\mathrm{bond}$ approaches $\frac12$ and
stays flat indicating a valence bond solid (VBS) phase with a rigid structure
where every second bond hosts a bonding state.
A more detailed inspection shows that in the VBS phase,
$E_\mathrm{GS}/E_\mathrm{bond}$ positively deviates from $\frac12$ with the
difference scaling as $K^6$. This energy gain can be understood within second order
perturbation theory as an effect of virtual processes where two neighboring $T$
pairs disappear to make space for an emerging middle $T$ pair 
(total amplitude is $K^3$) being an intermediate state.

Interestingly, the smooth evolution observed in Fig.~\ref{fig:chainK} suggests
a picture of the dilute triplon gas at $E_T\gg K$ being continuously connected
with the dense triplon state close to the QCP. It is further supported by a
gradual reduction of the spin gap closing at QCP Fig.~\ref{fig:chainK}(c) and
an exponential decay of dimer correlations [Fig.~\ref{fig:chainK}(d)] with the
decay length diverging at QCP when the VBS is formed.

In the Kitaev model, the spin gap separates the flux-free ground state from
the topological sector with two fluxes. Within this gap, excitations from the
flux-free sector carried by itinerant Majorana fermions can be found. In our
model, the spin gap shown in Fig.~\ref{fig:chainK}(c) separates the ground
state with all-even bond configuration and the sector with one odd bond that
is being flipped by the van Vleck moment operator. However, in contrast to the
Kitaev model, here the excitation to one-odd sector is lower than excitations
within all-even sector all the way up to QCP. In other words, no modes (e.g.
Majorana bands) are present within the spin-gap. At the QCP, the lowest
excitations merge, including also partly the lowest excitations to the other
sectors with more odd bonds. The special role of the QCP will be further
demonstrated in the next section -- an antiferromagnetic condensate will be
found to emanate from it and an intuitive picture of the critical excitations
near QCP will be inferred.


\subsection{Extension towards XY-chain}
\label{sec:chainKXY}

\begin{figure}
\begin{center}
\includegraphics[scale=1.00]{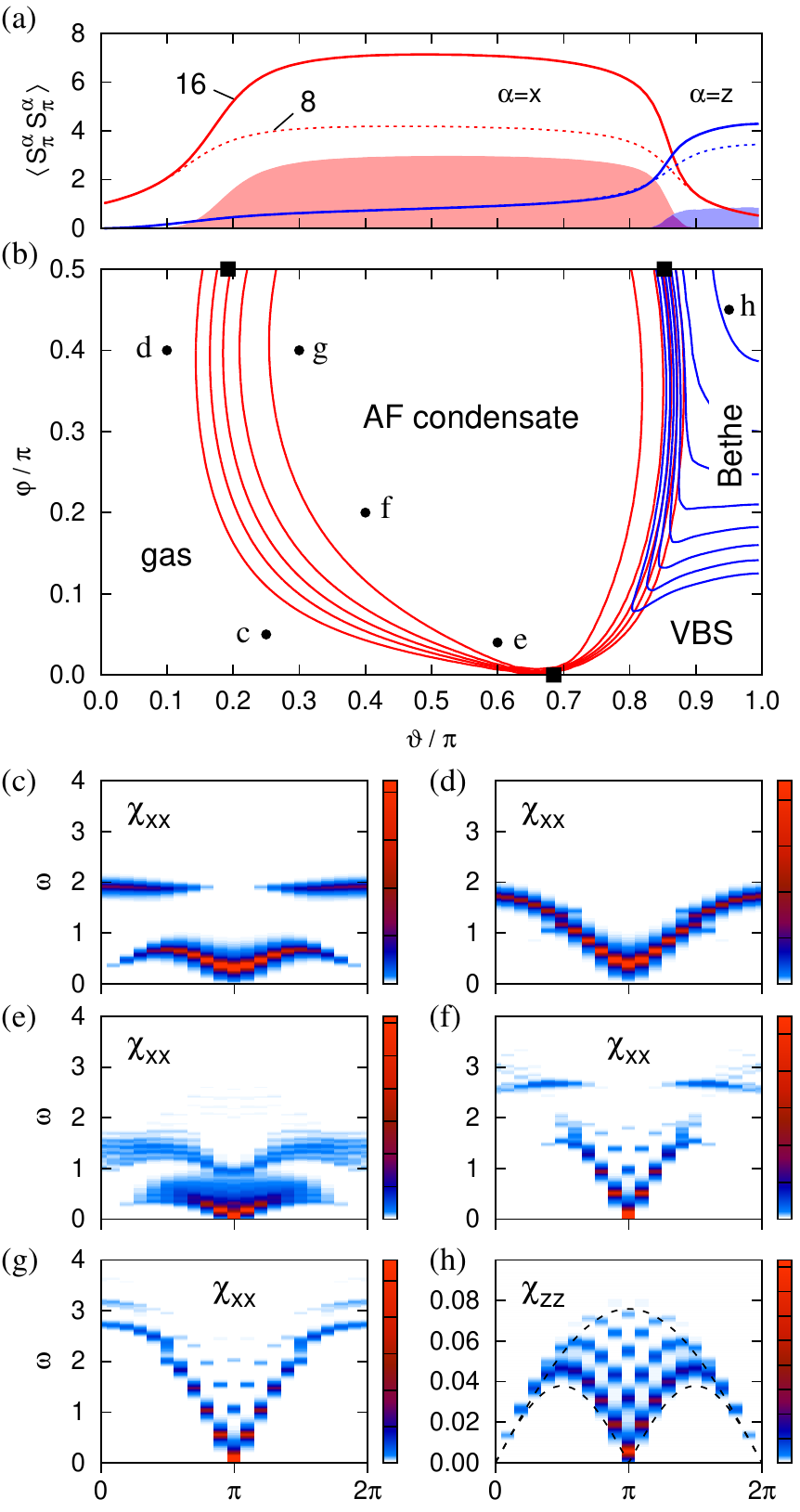}
\caption{
(a)~Correlations $\langle S^\alpha_q S^\alpha_{-q} \rangle$ with $q=\pi$ for
the chain model with $K=0$ parametrized as $E_T=\cos\vartheta$,
$J_\mathrm{XY}=\sin\vartheta$. Data for chains of the length $N=16$ and $N=8$
are shown. Their difference (shaded areas) represents the part of the
correlations scaling with the system size and could be used to estimate the
extent of the corresponding phases.
(b)~Approximate phase diagram of $E_T-K-J_\mathrm{XY}$ model parametrized as
$E_T=\cos\vartheta$, $K=\sin\vartheta\cos\varphi$,
$J_\mathrm{XY}=\sin\vartheta\sin\varphi$. The contours capture the difference
of the correlations $\langle S^\alpha_\pi S^\alpha_\pi \rangle$ between $N=16$
and $N=8$ chains. Red and blue lines are based on $S^x$ and $S^z$
correlations, respectively. Black squares are reference points from Langari
{\it et al.} \cite{Lan13} ($\varphi=\pi/2$ line) and from
Fig.~\ref{fig:chainK} ($\varphi=0$ line). The top line with $\varphi=\pi/2$
matches panel (a).
(c)--(h)~Imaginary part of $S^x$ or $S^z$ susceptibilities calculated for a
chain with $N=20$ sites at the selected parameter points marked in panel (b).
Panel (h) shows also the boundaries (dashed lines) of the excitation continuum
for the effective \mbox{spin-$\frac12$} Heisenberg chain.
}\label{fig:chainKJ}
\end{center}
\end{figure}

Having explored the Kitaev-like limit of the model on the chain, we now
consider its extension by $J_{XY}$-interaction introduced in
Eqs.~\eqref{eq:Ham1D} and \eqref{eq:Ham1Dspin1}. As our main interest is
the qualitative illustration of the concepts that appear in the honeycomb
case as well, we do not focus on the specifics that are related to
one-dimensionality but rather on the generic features that will be inherited
by the 2D lattice case. 

Let us start the discussion with the pure XY-limit where our model can be
related to \mbox{spin-$1$} XXZ chain with single-ion anisotropy for which
extensive studies are available \cite{Bot83,Sch86,Nij89,Che03,Lan13}, mainly
in the connection with the Haldane gap problem. Its phase diagram is quite
complex containing a number of phases depending on the
$J_\mathrm{Z}/J_\mathrm{XY}$ ratio and single-ion anisotropy $D$
[corresponding to our $E_T$ in Eq.~\eqref{eq:Ham1Dspin1}]: large-$D$ phase,
Haldane phase, two XY phases, the ferromagnetic phase, and the N\'eel phase
\cite{Che03}.  For the relevant $J_\mathrm{Z}=0$ case that matches to our
model, it shows two quantum critical points. The first one at $E_T \approx
0.34 J_\mathrm{XY}$ corresponds to the transition between the large-$D$ phase
and the Haldane phase and its precise determination requires the detection of
topological features of the Haldane phase such as the edge
\mbox{spin-$\frac12$} pair \cite{Che03}. The second transition to the N\'eel
phase occurs at $E_T\approx -2 J_\mathrm{XY}$ and in contrast to the first one
is easy to capture precisely \cite{Lan13}.

The XY-limit of our model is numerically studied in Fig.~\ref{fig:chainKJ}(a)
by means of spin correlations obtained by ED. Since the local conserved
quantities are lost when introducing the $J_\mathrm{XY}$ interaction, we are
now limited to a shorter chain length (at most $N=20$ sites) compared to the
previous paragraph. We employ both van Vleck moments [$S^x$ and $S^y$
components of the effective \mbox{spin-$1$} defined by
Eqs.~\eqref{eq:sTspin1a}, \eqref{eq:sTspin1b}] and triplon spin [$S^z$
component defined in Eq.~\eqref{eq:sTspin1c}]. The static correlators $\langle
S^\alpha_q S^\alpha_{-q} \rangle$ of their Fourier components
$S^\alpha_q=\sum_{l=1}^N S^\alpha_l \exp(-i q l)/\sqrt{N}$ at the
characteristic momentum $q=\pi$ are plotted for two different lengths of the
chain and subtracted.  The difference uncovers a correlation contribution
scaling with the system size (on top of a size-independent contribution) that
we regard as a signature of a particular phase.  Though oversimplified
compared to a full finite-size scaling analysis, this approach will later
provide a rough sketch of the phase diagram of the model in its entire
parameter space.

Figure~\ref{fig:chainKJ}(a) shows three regimes of the correlations for the XY
limit. The first one for $E_T\gg J_\mathrm{XY}$ corresponds to the triplon gas
with the correlations generated exclusively by triplon excitations. It is
quickly replaced at about $E_T \approx 1.5 J_\mathrm{XY}$ with a triplon
condensate characterized by antiferromagnetic (AF) correlations of van Vleck
moments 
$M_{1\alpha} \propto (s^\dagger T^{\phantom{\dagger}}_\alpha-T^\dagger_\alpha s)$.
An intuitive picture of the condensate can be based on a trial ground-state
wavefunction that explicitly mixes the condensed $T$ states into a ``pool'' of
$s$ states
\begin{equation}\label{eq:trial1D}
|\Psi\rangle = \prod_{l=1}^{N} \left( \sqrt{1-\rho}\, |s\rangle 
+ \sqrt{\rho}\, \;i\,\mathrm{e}^{i\pi l}\; |T\rangle \right)_l ,
\end{equation}
creating thus van Vleck moments. Here $|T\rangle$ stands for any normalized
combination of $|T_x\rangle$ and $|T_y\rangle$ and $\rho$ is the condensate
density, $0\leq \rho\leq 1$. At each site $l$ the hardcore condition $n_T\leq
1$ is obeyed.  While the wavefunction \eqref{eq:trial1D} is more appropriate
for the 2D case with static long-range AF order, it still captures the
transition between the regime of a triplon gas ($\rho=0$) and the condensate
with pronounced AF correlations ($\rho>0$) and gives a very crude estimate
$E_T=4J_\mathrm{XY}$ of the transition point. This is based on minimizing the
energy per site $E_T\rho-4J_\mathrm{XY}\rho(1-\rho)$ with respect to the
variational parameter $\rho$. Later in Sec.~\ref{sec:honeyfull}, we will
develop a quantitative mean-field treatment of the honeycomb model based on
the same type of condensation.

The second quantum phase transition (QPT) appearing at $E_T\approx -2
J_\mathrm{XY}$ is to a phase associated with the limit of large negative
$E_T$. At this transition, the energy gain from negative $E_T$ overcomes the
energy gain from correlated van Vleck moments and the system collapses to a
state full of $T_x$ and $T_y$ triplons leaving the triplon color as the only
active degree of freedom. The costly $s$ state can be integrated out leading
to an effective interaction among pseudospins-$\frac12$ describing the $T_x$,
$T_y$ doublets. In the isotropic XY case under consideration, the resulting
effective model valid for $J_\mathrm{XY}\ll -E_T$ is simply a
\mbox{spin-$\frac12$} Heisenberg chain with the exchange parameter
$J_\mathrm{eff}=J_\mathrm{XY}^2/|E_T|$. It can be obtained by removing $s$ via
second order perturbation theory and introducing the sublattice dependent
mapping
\begin{align}
|T_x\rangle &\rightarrow |\!\uparrow\rangle, &
|T_y\rangle &\rightarrow |\!\downarrow\rangle &\text{(even sites)},\\
|T_x\rangle &\rightarrow |\!\downarrow\rangle, &
|T_x\rangle &\rightarrow -|\!\uparrow\rangle &\text{(odd sites)} .
\end{align}
Because of the connection to the exactly solvable \mbox{spin-$\frac12$}
Heisenberg chain, hereafter we call the corresponding phase the Bethe phase
(N\'eel phase in the context of \mbox{spin-$1$} XXZ chains).

It has to be noted that while our second QPT for negative $E_T$ well
corresponds to the reference data by Chen {\em et al.} \cite{Che03}, the first
change of the regime occurs for much smaller $J_\mathrm{XY}$ in our data than
in Ref.~\cite{Che03}. On the other hand, we obtain a good agreement with the
quantum renormalization group (QRG) and ED study \cite{Lan13} of the
ground-state fidelity.
This may be interpreted within the triplon condensation picture as follows. 
The true Haldane phase appears around $E_T \approx 0.34 J_\mathrm{XY}$ but
this QPT is preceded much earlier by our ``transition'' associated with the
onset of van Vleck correlations and corresponding to the emergence of a
triplon (quasi-)condensate. Such a change in the ground-state structure is
also reflected in the ground-state fidelity inspected in Ref.~\cite{Lan13}.
While probably not a real QPT, it is a crossover determined by the energy
balance of the triplon cost and the energy gain due to a formation of
correlated van Vleck moments. In non-frustrated situations, this energy
balance shall lead to a crossover/transition at similar $J/E_T$ ratios,
depending mainly on the connectivity of the particular lattice. Therefore, the
apparent discrepancy is not essential because in the 2D honeycomb case, the
emergence of the condensate will correspond to establishing a real long-range
AF order of van Vleck moments.

After discussing both the Kitaev-like limit explored in Sec.~\ref{sec:chainK}
and the XY-limit in the above, we now extend the correlation-based approach to
the full $E_T-K-J_\mathrm{XY}$ model to obtain a sketch of the phase
diagram presented in Fig.~\ref{fig:chainKJ}(b).

The topology of the phase diagram follows from the features already met above
when inspecting the limiting cases. Most of the phase diagram, in particular
all of its physically sensible part ($E_T>0$), is taken up by the competition
of the triplon gas and the triplon condensate with AF correlations of van
Vleck moments -- components $S^x$, $S^y$ of the effective \mbox{spin-$1$}. The
crossover is more and more delayed when going from the XY-limit
($\varphi/\pi=0.5$) to the Kitaev-like one ($\varphi=0$). This is easily
understood by an increasing frustration in this direction and thus a smaller
gain from creating correlated van Vleck moments. A larger interaction strength
is thus needed to overcome the $E_T$ cost. The remaining phases are restricted
to the area of large negative $E_T$.  Depending on the balance between $K$ and
$J_\mathrm{XY}$, the system selects either the VBS phase with every second
bond essentially inactivated, or the Bethe phase linked to the hidden
effective \mbox{spin-$\frac12$} model -- a Heisenberg chain -- and revealed by
AF correlations of $S^z$ components of the effective \mbox{spin-$1$}.

To complement the phase diagram, Figs.~\ref{fig:chainKJ}(c)--(h) present
dynamical correlations 
$\chi_{\alpha\alpha}(q,\omega)=\langle S^\alpha_q S^\alpha_{-q} \rangle_\omega$,
i.e. spin susceptibility associated with the effective \mbox{spin-$1$},
calculated for several points in the phase diagram. The gapped van Vleck
susceptibility $\chi_{xx}=\chi_{yy}$ in Fig.~\ref{fig:chainKJ}(c),(d) for the
triplon gas phase shows the difference between the local-like response
consisting of two flat parts in the Kitaev-like limit
[Fig.~\ref{fig:chainKJ}(c)] and (almost) continuous dispersion at large
$J_\mathrm{XY}$ [Fig.~\ref{fig:chainKJ}(d)]. Similarly,
Figs.~\ref{fig:chainKJ}(e)-(g) capture the evolution from the Kitaev-like to
the XY-limit response of the AF condensate. Here the low-energy part is
dominated by an intense linear mode centered at the AF wavevector $q=\pi$.
Extrapolation of data up to $N=20$ suggests gapless response inside the AF
condensate region, within the precision limited by finite-size effects that
are pronounced mainly in the transition region.  Finally, inspecting the
susceptibility $\chi_{zz}$ for a point deep inside the Bethe phase, we notice
that the dynamical response clearly reveals the hidden \mbox{spin-$\frac12$}
Heisenberg chain. For example, its excitation continuum perfectly matches the
expected analytical boundaries obtained using
$J_\mathrm{eff}=J_\mathrm{XY}^2/|E_T|$, see the dashed lines in
Fig.~\ref{fig:chainKJ}(h).

An interesting feature is the ``emanation'' of the AF condensate from the QCP
of the Kitaev-like model. Around that point, indicated by a black square on
$\varphi=0$ line of Fig.~\ref{fig:chainKJ}(b), the triplon gas consists of
bonding states that are about to form VBS, while for the AF condensate above
QCP we expect the state of the type \eqref{eq:trial1D}.  A naive picture of
the link between the two states that is related to low-energy van Vleck
excitations observed in Fig.~\ref{fig:chainK}(c) can be constructed as
follows: For simplicity, let us consider mixing of $s$ and $T$ states in 1:1
ratio and ignore the triplon color.  Adopting the condensate wavefunction
\eqref{eq:trial1D} with $\rho=\frac12$, at two neighboring sites $l,l+1$ we
have a state proportional to
\begin{multline}\label{eq:condbond}
|\mathrm{cond}\rangle \propto
(|s\rangle+i|T\rangle)_l\otimes(|s\rangle-i|T\rangle)_{l+1} \\ 
= [\,|ss\rangle+|TT\rangle-i(|sT\rangle-|Ts\rangle)\,]_{l,l+1}.
\end{multline}
On the other hand, the bonding state with 1:1 mixing is 
$|B\rangle \propto |ss\rangle + |TT\rangle$. Applying the van Vleck operator 
$M_q \propto -i \sum_{l=1}^N \mathrm{e}^{-i q l}(
s^\dagger_l T^{\phantom{\dagger}}_l - T^\dagger_l s^{\phantom{\dagger}}_l)$
with the critical $q=\pi$ on $|B\rangle$, we obtain
$-2i(|sT\rangle-|Ts\rangle)$ which is exactly the missing part to get bond
state $|\mathrm{cond}\rangle$ of \eqref{eq:condbond}.  The presence of
low-energy van Vleck excitations that become gapless at the QCP of the
Kitaev-like limit therefore makes the ``pool'' of bonding states susceptible
to the formation of the AF condensate of the type \eqref{eq:trial1D} and this
condensate is indeed formed once $J_\mathrm{XY}$ is added. 



\section{Honeycomb lattice case}
\label{sec:honey}

With the basic physical features of our model being illustrated by the 1D
simplified case, we now focus on its original version on the honeycomb
lattice. While the 2D lattice and one more degree of freedom bring an
increased complexity compared to what discussed in Sec.~\ref{sec:chain}, the
overall behavior will turn out to be rather similar.

\subsection{The case of pure Kitaev-like interaction}
\label{sec:honeyK}

\begin{figure}
\begin{center}
\includegraphics[scale=0.99]{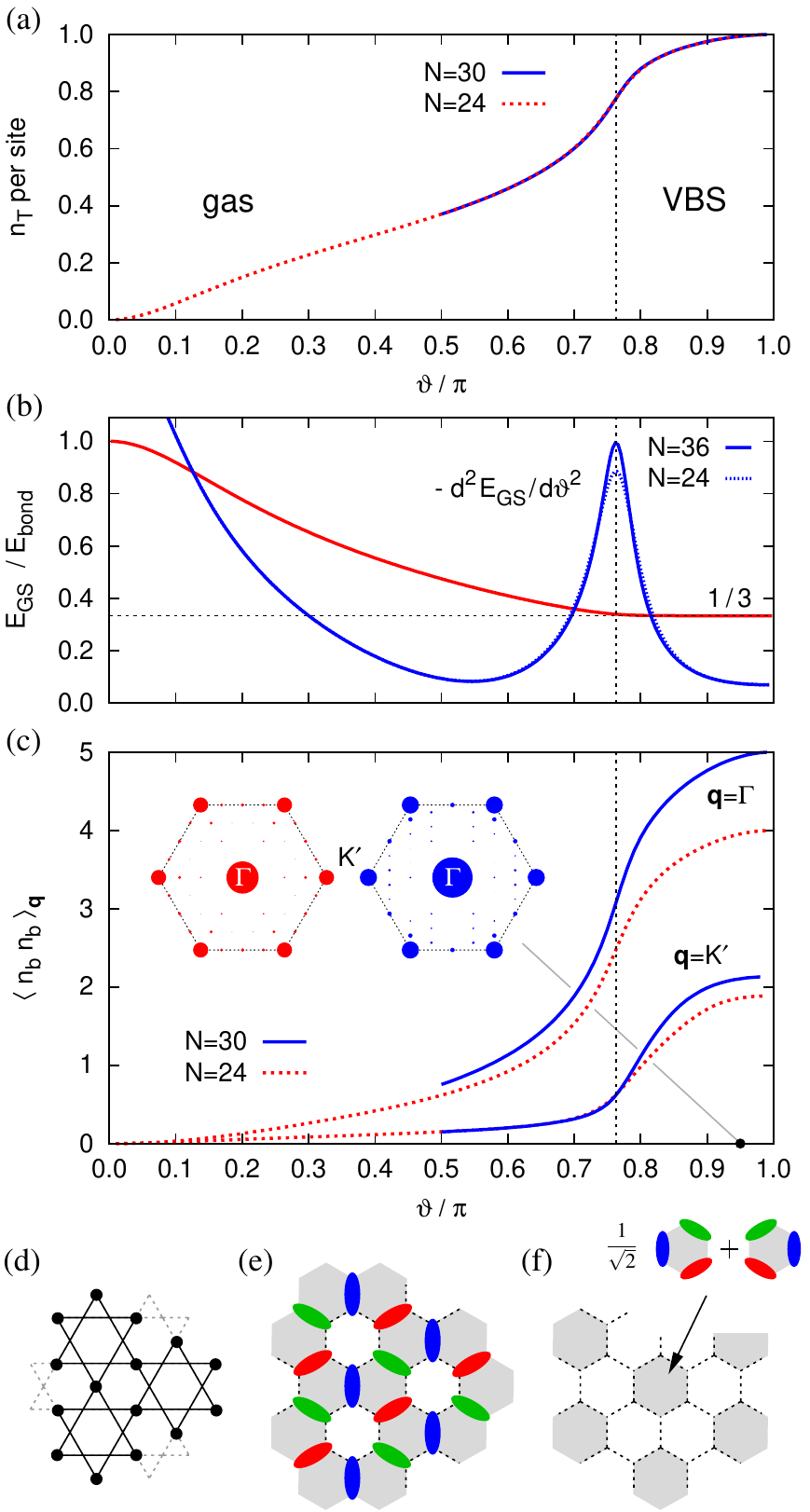}
\caption{
(a)~Occupation of the triplet state within the honeycomb model parametrized as
$E_T=\cos\vartheta$, $K=\sin\vartheta$. Presented for a hexagonal $24$-site
cluster and rectangular $30$-site cluster.
(b)~Ground-state energy per bond $E_\mathrm{GS}$ measured by bonding-state
energy $E_\mathrm{bond}$ (red) and the second derivative of $E_\mathrm{GS}$
with respect to $\vartheta$ revealing the quantum critical point (blue).
(c)~Correlations $\langle n_b n_b\rangle_{\vc q}$ of the bosons $b$ on the
dual kagom\'e lattice. Size-dependent correlations at the characteristic
vector $\vc q=K'$ detect the VBS state. The insets show the correlations  at
$\vartheta=0.95\pi$ plotted in the extended Brillouin zone of the kagom\'e
lattice at all momenta accessible when using the $24$- and $30$-site clusters,
respectively. 
(d)~Static pattern of the $b$ bosons on the dual lattice suggested by their
reciprocal-space correlations.
(e)~Corresponding arrangement of $T$-dimers in the original representation.
Red, green, and blue ellipses represent bonding states involving $T_x$, $T_y$,
and $T_z$, respectively. The shaded hexagons indicate flippable plaquettes.
(f) The actual plaquette order in VBS phase. The shaded hexagons indicate
resonating plaquettes.
}\label{fig:honeyK}
\end{center}
\end{figure}

This similarity is seen already in Fig.~\ref{fig:honeyK}(a),(b) which is a
direct analogy of Fig.~\ref{fig:chainK}(a),(b) capturing the competition
between the gas and VBS phase in the Kitaev-like limit of the model. Again a
single QCP is detected, now with a position shifted to a smaller $K/|E_T|$
ratio in the negative $E_T$ region. The reduction of the VBS phase is a
consequence of the lattice connectivity -- the VBS state can only host a
bonding state on one third of the bonds compared to one half in the chain
case, leading to a less competitive energy gain. 
The formation of VBS consisting of maximum geometrically possible number of
dimers (bonding states), i.e. $N_\mathrm{bond}/3$, is seen also in the GS
energy per bond measured by $E_\mathrm{bond}$. This quantity stays close to
$\frac13$ and there is again a small positive deviation scaling as $K^6$ that
indicates residual interactions among the dimers. They establish a specific
dimer arrangement that we detect in Fig.~\ref{fig:honeyK}(c) using dual
$b$-boson representation on the kagom\'e lattice as described in
Sec.~\ref{sec:dualmap} (note that all the bond parities are even in the case
inspected). The VBS phase is marked by size-dependent reciprocal-space
correlations of the $b$-boson density $n_b$ at the characteristic momenta $\vc
q=K'$ lying in the corners of the extended Brillouin zone of the kagom\'e
lattice. This suggests the real-space pattern shown in
Fig.~\ref{fig:honeyK}(d) that is also the most probable $b$ configuration in
the ground state. Translating it into the $T$-dimer picture, we obtain the
pattern in Fig.~\ref{fig:honeyK}(e) which maximizes the number of plaquettes
carrying three dimers. However, the true structure of the ground state is more
complicated and requires the following deeper analysis.

In fact, working in the basis of maximum dimer coverings, an effective quantum
dimer model (QDM) with $\mathcal{O}(K^6)$ interactions can be formulated.
This QDM ``lives'' in the all-even parity sector and captures both the ground
state and the lowest excitations.
Leaving the details aside, we note that the dimer model involves two kinds of
interactions: First, there is an energy gain $\propto K^6$ from dimer-dimer
bonds which is constant for all the coverings and which was already noticed in
the 1D case with just two trivial dimer coverings. The second contribution,
being the actual driving force stabilizing the VBS pattern, is enabled by the
geometry of the honeycomb lattice and corresponds to hexagonal plaquette flips 
with an amplitude $\propto K^6$. They are captured by the QDM Hamiltonian
\begin{equation}\label{eq:HQDM}
\mathcal{H}_\mathrm{QDM} = -t \sum_\mathrm{plaquettes} 
\left(
\left|       \raisebox{-6pt}{\includegraphics[height=18pt]{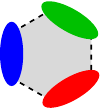}} \right\rangle
\left\langle \raisebox{-6pt}{\includegraphics[height=18pt]{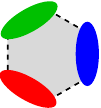}} \right|
+ \mathrm{H.c.} 
\right) ,
\end{equation}
with $t=3K^6/16|E_T|^5$, i.e. by the kinetic term of the Rokhsar--Kivelson
(RK) model for the honeycomb lattice \cite{Rok88,Moe01}. Using the connection
to this well-studied model we can fix the type of order in the VBS phase now.
The phase diagram of the honeycomb RK model, depending on the ratio of the
flippable plaquette energy cost $V$ and the flipping amplitude $t$, was
precisely determined by Monte Carlo simulations \cite{Moe01,Sch15,Sch17}. Our
\mbox{$V/t=0$} case falls into the interval between $(V/t)_c \approx -0.23$
\cite{Sch17} and the RK point $V/t=1$ where the honeycomb RK model supports a
triangular covering by resonating plaquettes (``plaquette'' order) depicted
approximately in Fig.~\ref{fig:honeyK}(f) which is the true VBS order for our
model. The anticipated ``columnar'' order of Fig.~\ref{fig:honeyK}(e) only
appears below the (rather close) critical point $(V/t)_c$ of the RK model
where the sufficiently large negative potential energy of the plaquettes wins.

Finally, we conclude the comparison to the Kitaev-like 1D model by a remark
that the spin gap behavior for the honeycomb lattice strongly resembles that
of the 1D chain case [see Fig.~\ref{fig:chainK}(c)], i.e. the spin gap
gradually closes as we approach the QCP from both the gas as well as VBS
phases.


\subsection{Full honeycomb model}
\label{sec:honeyfull}

\begin{figure}
\begin{center}
\includegraphics[scale=0.94]{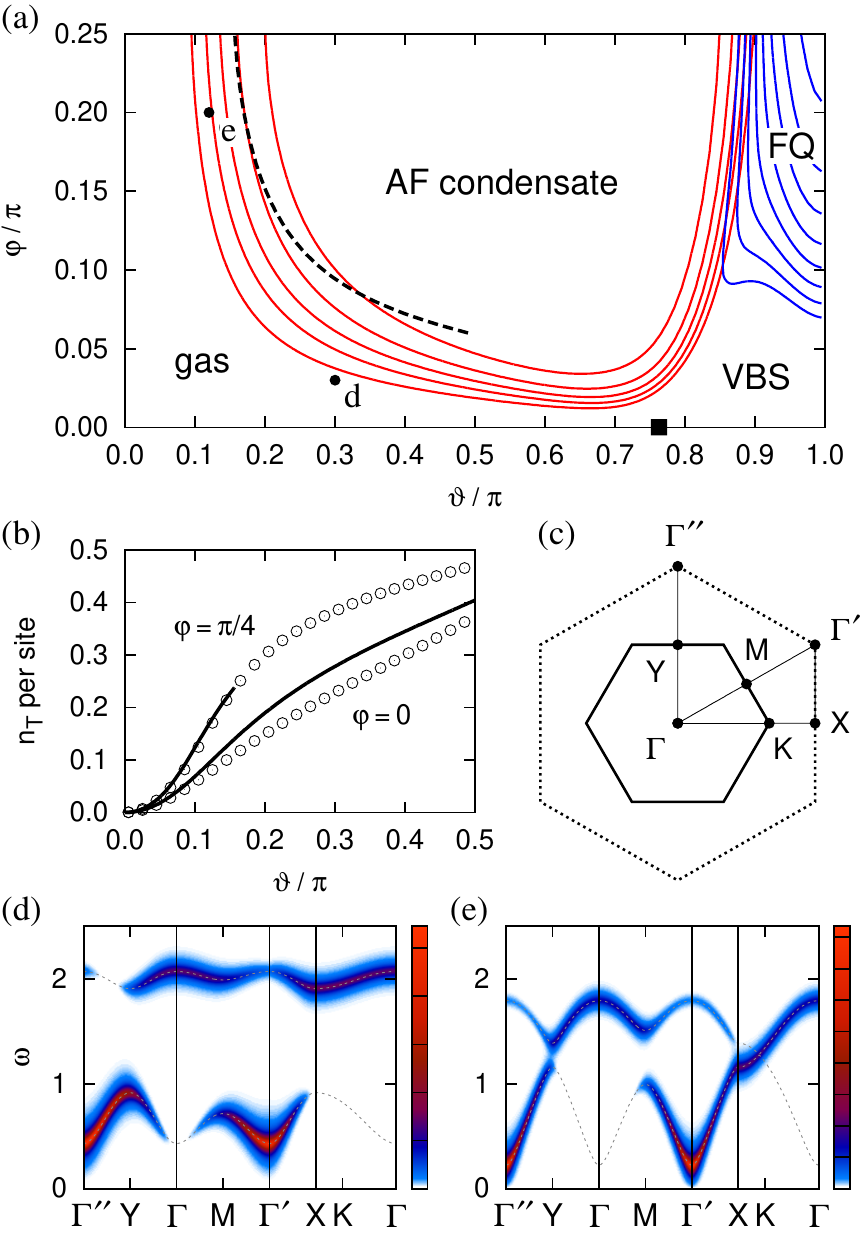}
\caption{
(a)~Approximate phase diagram of $E_T-K-\tilde{K}$ model up to
$\tilde{K}\!=\!K$ estimated as in Fig.~\ref{fig:chainKJ}(b) via size-dependent
characteristic correlations. Hexagonal $6$-site cluster and rectangular
$12$-site cluster were used. The model is parametrized as:
$E_T=\cos\vartheta$, $K=\sin\vartheta\cos\varphi$,
$\tilde{K}=\sin\vartheta\sin\varphi$.  Red lines show contours of the increase
of the van Vleck correlations $\langle S^x_{\vc q} S^x_{-\vc q}\rangle$ at the
antiferromagnetic (AF) momentum $\vc q=\Gamma'$ when going from $6$- to
$12$-site cluster. The correlations detecting the ferroquadrupolar (FQ) phase
(blue contours) were estimated as described in the text.  Black square is the
extrapolated position of the QCP for pure Kitaev-like ($K$ only) model [see
Fig.~\ref{fig:honeyK}(b)]. The dashed line indicates the transition between
the gas phase and AF condensate obtained using Gutzwiller-type treatment of
the hardcore constraint.
(b)~Comparison of mean-field (solid line) and ED results (circles, 12-site
cluster) for $n_T$ at the top ($\varphi=\pi/4$) and bottom line ($\varphi=0$)
of the phase diagram.
(c)~Brillouin zone of the honeycomb lattice (solid) and completed triangular
lattice (dotted) with indicated special points and the path used in the next
panels.
(d),(e)~Imaginary part of van Vleck spin susceptibility $\chi_{zz}(\vc
q,\omega)$ calculated in the mean-field approximation for the two points
marked by d and e in panel (a) and artificially broadened. The dotted lines
indicate the dispersions of Eq.~\eqref{eq:dispBA} with $\alpha=z$.
}\label{fig:honeyKKt}
\end{center}
\end{figure}

In this section we explore the phase diagram and to a limited extent also the
excitations of the full model of Eq.~\eqref{eq:Ham} containing both the
Kitaev-like interaction $K$ and the complementary one $\tilde{K}$. We do not
go up to the dominant $\tilde{K}$ regime characterized by strongly interacting
quasi-one-dimensional condensates hosted by zigzag chains in the honeycomb
lattice \cite{Kha13}. Instead, similarly to the chain case, we interpolate
between the Kitaev-like limit and the isotropic limit $K=\tilde{K}$, being
both positive as derived. Phase diagram for arbitrary $\tilde{K}/K$ ratio and
positive $E_T>0$ sector can be found in Ref.~\cite{Ani19}. 


\subsubsection{Phase diagram}

Fig.~\ref{fig:honeyKKt}(a) presents the phase diagram of the $E_T-K-\tilde{K}$
model estimated by the method of Sec.~\ref{sec:chainKXY}, i.e. by tracking the
cluster-size-dependent correlations characteristic to the individual phases.
Due to a larger local basis of four states, lack of symmetries, and many
points to be analyzed, we had to resort to a combination of small $6$-site and
$12$-site clusters. This makes the phase diagram rather semi-quantitative as
seen for example by comparing the position of the QCP in the Kitaev-like limit
obtained for much bigger clusters in Sec.~\ref{sec:honeyK}. Nevertheless, four
phases in an arrangement resembling that of Fig.~\ref{fig:chainKJ}(b) can be
identified. Two of them -- gas and VBS phases -- were already encountered in
the previous Sec.~\ref{sec:honeyK}. 

The left part of the phase diagram ($E_T>0$) is a playground for a competition
between the AF correlations and Kitaev-like frustration of the interactions.
In the isotropic $K\!=\!\tilde{K}$ limit [top line in
Fig.~\ref{fig:honeyKKt}(a)] where the bond interactions handle all the triplon
colors equally, the situation is analogous to the square-lattice case
discussed in the context of \mbox{Ca$_2$RuO$_4$} \cite{Kha13,Jai17,Sou17}.
Since the honeycomb lattice is not geometrically frustrated, it can easily
host an antiferromagnetic phase associated with a bipartite condensation of
triplons captured by a wavefunction similar to that of Eq.~\eqref{eq:trial1D}.
This AF phase with ``soft'' (i.e. far from saturated) van Vleck moments $\vc
M_1$ actually takes the largest portion of the entire phase diagram in
Fig.~\ref{fig:honeyKKt}(a). Going down to the region with strong Kitaev-like
anisotropy of the interactions and the resulting frustration, the AF phase
gets largely suppressed. 
One of the highlights of the soft-moment magnetism based on spin-orbit triplon
condensation is the presence of both transverse magnon modes as well as an
intense amplitude mode (dubbed ``Higgs mode'' in this context), that have been
observed experimentally in Ca$_2$RuO$_4$ \cite{Jai17,Sou17}. It is an
interesting and non-trivial problem for a future study to analyze the effect
of the Kitaev-like frustration on such magnetic excitation spectra. Later in
Sec.~\ref{sec:dyngutzw}, we only make an attempt to address the gas phase
close to the AF phase boundary, by inspecting the excitation spectrum close to
the point where the condensate is formed.

Focusing now on the large negative $E_T$ limit, we can again notice
similarities to the 1D chain case of Sec.~\ref{sec:chainKXY}. In this limit
the singlets $s$ can be integrated out leaving us with an effective spin-1
model where the spin now coincides with the triplon spin $-i(\vc
T^\dagger\times\vc T)$. The nature of this model changes with the
$\tilde{K}/K$ ratio. For the strongly anisotropic $K$-only case, one
formally obtains a biquadratic Kitaev-like model as a leading term:
\begin{equation}\label{eq:HbqK}
\mathcal{H}_\mathrm{bqK} = 
-\frac{K^2}{2|E_T|}\, \sum_{\langle ij\rangle} (S^\alpha_i S^\alpha_j)^2 ,
\end{equation}
with the active component $\alpha\in\{x,y,z\}$ given by the bond direction as
before. Compared to the usual bilinear spin-one Kitaev model (see, e.g., 
Ref.~\cite{Min19} and references therein), the behavior of the biquadratic model 
is rather trivial. In the original language, it simply counts the number of
proper-color $T$-dimers and associates an energy gain ${K^2}/2|E_T|$ with each
of them. This selects a large number of degenerate $T$-dimer coverings as the
model ground state. At this level of approximation which misses the
$\mathcal{O}(K^6)$ interactions, the true VBS ground state cannot be resolved.
Furthermore, the low-lying excitations with the energies $\propto K^6$ are not
captured. Hence, in the Kitaev-like limit, spin-1 is not a suitable
elementary object and bonding-state dimers should be used instead, leading to
the effective quantum dimer model which we extensively discussed in 
Sec.~\ref{sec:honeyK}. 

In contrast to that, the isotropic limit $K\!=\!\tilde{K}$ can be expected to
be adequately captured by a simple isotropic spin-1 model. Indeed, the
preference of the total-singlet $T$-pairs that may virtually transform into
$s$-pairs and back gives rise to biquadratic interaction described by the
effective Hamiltonian
\begin{equation}\label{eq:Hbq}
\mathcal{H}_\mathrm{bq} = 
-\frac{K^2}{2|E_T|}\, \sum_{\langle ij\rangle} (\vc S_i\vc S_j)^2
\end{equation}
at the isotropic point $K\!=\!\tilde{K}$. Model of this kind is a special case
of bilinear-biquadratic spin-1 models that were thoroughly explored for
various lattices. On a non-bipartite, geometrically frustrated triangular
lattice \cite{Lau06,Tsu06} its ground state shows a ferroquadrupolar (FQ)
order \cite{Che73,Har02,Iva03} which does not break time-reversal symmetry but
introduces a preferential plane in spin space where the spins can be found
with a higher probability. For a non-frustrated lattice such as square
\cite{Har02,Tot12} and honeycomb \cite{Lee12,Zha12,Cor13} the AF phase is more
competitive and the biquadratic model is just on the verge of the FQ and AF
order. This type of order, labeled as FQ in Fig.~\ref{fig:honeyKKt}(a) for
simplicity, therefore replaces the Bethe phase of the 1D chain [compare
Fig.~\ref{fig:chainKJ}(b)]. Note that here we refer to the AF phase of triplon
spins $-i(\vc T^\dagger\times\vc T)$; this should not be confused with the
neighboring AF phase of correlated van Vleck moments $\vc M_1$ which reside on
the $J=0\leftrightarrow 1$ transitions. 
The detection of the ``edge-case'' FQ/AF order is somewhat complicated, also
due to the incompatible geometry of the two clusters ($6$-site hexagon and
$12$-site rectangle) that we use to check the size-scaling of the
characteristic correlations. To this end, as the characteristic quantity we
take the contribution to FQ correlations $\langle Q_{\vc q} Q_{-\vc q}
\rangle$ with $\vc q=0$ that is carried by triplon spins at AF momentum $\vc
q=\Gamma'$. More explicitly, we decompose the various quadrupole operators $Q$
containing $S^\alpha S^\beta$ terms (see e.g. Ref.~\cite{Lau06} for their
explicit forms) into a momentum sum
$Q_{\vc q} \sim \sum_{\vc q'} S_{\vc q-\vc q'}^\alpha S_{\vc q'}^\beta$
and evaluate the four-spin correlators of the type
$\langle S_{\vc q-\vc q'}^\alpha S_{\vc q'}^\beta 
S_{-\vc q-\vc q''}^\gamma S_{\vc q''}^\delta \rangle$ constituting 
$\langle Q_{\vc q} Q_{-\vc q} \rangle$.  The contribution with all the momenta
being equal to the AF one is found to dominate and behave well at the
reference $K=\tilde{K}\ll |E_T|$ point described by $\mathcal{H}_\mathrm{bq}$
of Eq.~\eqref{eq:Hbq}.  The resulting correlations obtained as a difference
between 12-site and 6-site cluster are shown in Fig.~\ref{fig:honeyKKt}(a).
They suggest that FQ/AF phase extends slightly further from its reference
point than the VBS one, though this result may be potentially biased as the
small clusters can not properly accommodate the VBS state. As a general remark
on biquadratic spin models such as \eqref{eq:Hbq}, it is worth noticing that
while they are typically very weak in conventional spin systems, they emerge
naturally in singlet-triplet level systems; see yet another example in
Ref.~\cite{Cha13}. 


\subsubsection{Dynamical Gutzwiller treatment and excitations}
\label{sec:dyngutzw}

As argued in Sec.~\ref{sec:spincorr}, the Kitaev-like anisotropy of the
interactions should manifest itself by the localized nature of the dynamic
spin response which translates to the flat dispersions of the spin
excitations. Having now covered the entire interval from the Kitaev-like limit
to the isotropic limit, one might wonder about the corresponding evolution of
the spin excitations. Due to the limited cluster size, ED calculations do not
provide a sufficient resolution to study such effects. 

Here we adopt instead a dynamical Gutzwiller treatment combined with
selfconsistent mean-field approximation formulated for the triplon gas phase.
Besides the excitation spectrum, this enables us to obtain the gas/AF
transition point as a function of non-Kitaev term $\tilde K$. The derivation
starts with a replacement of $s$ implicitly contained in Eq.~\eqref{eq:Ham} by
the operator $\sqrt{1-n_T}$ which dynamically accounts for the triplon
hardcore constraint. The resulting Hamiltonian is expanded and a mean-field
decoupling is applied leading to the quadratic Hamiltonian
\begin{multline}\label{eq:HamMF}
  \mathcal{H}_\mathrm{MF} = \sum_i (E_T+\Lambda)\, n_{T_i} \\
  + \sum_{{\langle ij\rangle}_\alpha} \langle 1- n_T\rangle
[K\mathcal{O}_{ij}^{\alpha} 
+\tilde{K}(\mathcal{O}_{ij}^{\bar{\alpha}}
          +\mathcal{O}_{ij}^{\bar{\bar{\alpha}} }) ] , 
\end{multline}
where all the $s$, $s^\dagger$ operators in $\mathcal{O}$ are left out.  At
this point we have already relaxed the hardcore constraint.  Two effects of
the hardcore nature of the triplons got captured at the level of
$\mathcal{H}_\mathrm{MF}$: (i) effective triplon cost is increased by an
energy  
\begin{equation}\label{eq:chempot}
\Lambda = -\frac12 \sum_{\delta_\alpha} \bigl[
K\langle\mathcal{O}_{i,i+\delta}^{\alpha}\rangle
+\tilde{K}(\langle\mathcal{O}_{i,i+\delta}^{\bar{\alpha}}\rangle
          +\langle\mathcal{O}_{i,i+\delta}^{\bar{\bar{\alpha}} }\rangle)
\bigr]
\end{equation}
($\delta$ runs through all nearest-neighbors), and (ii) the interactions $K$,
$\tilde{K}$ are reduced by a factor $1-\langle n_T\rangle$ that embodies the
probability of another triplon blocking the interaction process on the
particular bond.  After a conversion to momentum space we obtain
\begin{multline}\label{eq:HamMFk}
\mathcal{H}_\mathrm{MF} = \sum_{\vc k\alpha} E\, \bigl(
\alpha_{1\vc k}^\dagger \alpha_{1\vc k}^{\phantom{\dagger}} +
\alpha_{2\vc k}^\dagger \alpha_{2\vc k}^{\phantom{\dagger}} 
\bigr) \\
+\Bigl[\kappa_{\alpha\vc k} \bigl(
\alpha_{1\vc k}^\dagger \alpha_{2\vc k}^{\phantom{\dagger}} -
\alpha_{1,-\vc k}^{\phantom{\dagger}} \alpha_{2\vc k}^{\phantom{\dagger}}
\bigr) 
+\mathrm{H.c.}\Bigr],
\end{multline}
where the unconstrained triplons are labeled by their color $\alpha=x,y,z$ and
the index $1$ or $2$ referring to the two sites in the unit cell of the
honeycomb lattice. It is convenient to choose the unit cell for triplons of
color $\alpha$ as a bond of direction $\alpha$. In this convention, the
on-site triplon energy $E=E_T+\Lambda$ entering Eq.~\eqref{eq:HamMFk} is
complemented by the interaction term $\kappa_{\alpha\vc k}=\langle 1-
n_T\rangle (K+\tilde{K}\eta_{\alpha \vc k})$ with the formfactor $\eta_{z\vc k}$
given by $\eta_{z\vc k}=2\cos ( \frac{\sqrt3}2 k_x) \exp (-i\frac32 k_y)$,
and $\eta_{x\vc k}$ and $\eta_{y\vc k}$ being simply rotated by multiples of
$2\pi/3$. Note that $\kappa_{\alpha\vc k}$ depends on momentum $k$ via
non-Kitaev $\tilde K$-term only.

The $4\times 4$ problem contained in Eq.~\eqref{eq:HamMFk} can be diagonalized 
explicitly and yields the dispersions of the bosonic quasiparticles
\begin{align}\label{eq:dispBA}
\omega_{B\alpha\vc k} &= \sqrt{E(E-2|\kappa_{\alpha\vc k}|)}, \notag \\
\quad 
\omega_{A\alpha\vc k} &= \sqrt{E(E+2|\kappa_{\alpha\vc k}|)}.
\end{align}
The averages entering all the equations starting from Eq.~\eqref{eq:HamMF}
have to be calculated self consistently via
\begin{align}
\langle n_T \rangle &= \frac1{4}\sum_{\vc k\alpha} \left(
\frac{E-|\kappa_{\alpha\vc k}|}{\omega_{B\alpha\vc k}}+
\frac{E+|\kappa_{\alpha\vc k}|}{\omega_{A\alpha\vc k}}
-2\right) , \\
\Lambda &= \frac1{4}\sum_{\vc k\alpha} |K+\tilde{K}\eta_{\alpha\vc k}|
\left(\frac{E}{\omega_{B\alpha\vc k}}-\frac{E}{\omega_{A\alpha\vc k}}\right) .
\end{align}

The above approach is applicable through the entire gas phase where the
excitations are found to be gapped ($\omega_{B\alpha\vc k}$ and
$\omega_{A\alpha\vc k}>0$).  Once the lower-energy $\omega_{B\alpha\vc k}$
touches zero at some point of the Brillouin zone, the triplon condensation
occurs with the condensate structure being similar to the one of
Eq.~\eqref{eq:trial1D}.  The corresponding condition $E=2|\kappa_{\alpha\vc
k}|$ is first met at $\vc k=0$ which results in the following equation:
\begin{equation}
E_T+\Lambda = 2 \langle 1- n_T\rangle (K+2\tilde{K}) 
\end{equation}
determining the points where AF condensate starts to form. The gas/AF phase
boundary obtained this way is presented as a dashed line in
Fig.~\ref{fig:honeyKKt}(a). It shows a good overall agreement with the
estimate by ED, correctly capturing the physical trend of a delayed
condensation when the frustration increases approaching the Kitaev-like limit.
As expected, the best agreement is obtained near the isotropic limit which is
also illustrated in Fig.~\ref{fig:honeyKKt}(b) where the isotropic-limit data
($\varphi=\pi/4$) of the selfconsistent $\langle n_T\rangle$ perfectly match
the ED values. In the Kitaev-like limit ($\varphi=0$), the deviation is
already significant but still acceptable for our semi-quantitative analysis.

With an adequate description of the excitations in the gas phase at hand, we
are now ready to inspect an analogy to Fig.~\ref{fig:chainKJ}(c),(d) presenting
the dynamical spin susceptibility for the gas phase of the 1D chain model. To 
this end we express the Fourier component of the van Vleck moment operator 
$\propto -i( s^\dagger T^{\phantom{\dagger}}_\alpha-T^\dagger_\alpha s)$
in terms of the unconstrained triplons as
\begin{equation}
S^\alpha_{\vc q} \propto -i \left[
\alpha_{1\vc q}^{\phantom{\dagger}} - \alpha_{1,-\vc q}^\dagger 
+ 
( \alpha_{2\vc q}^{\phantom{\dagger}} - \alpha_{2,-\vc q}^\dagger )
\mathrm{e}^{-i \vc q \cdot \vc \delta_\alpha}
\right] ,
\end{equation}
where $\vc \delta_\alpha$ is a unit vector in the direction of $\alpha$ bonds.
Next, we use the eigenspectrum of $\mathcal{H}_\mathrm{MF}$ to find the dynamic 
correlation function $\chi_{\alpha\alpha}(\vc q,\omega)=
\langle S^\alpha_{\vc q} S^\alpha_{-\vc q} \rangle_{\omega}$
shown in Fig.~\ref{fig:honeyKKt}(d),(e) for two selected points in the phase
diagram. 

Similarly to the chain case, the vicinity of the Kitaev-like limit
[Fig.~\ref{fig:honeyKKt}(d)] is characterized by flat dispersion of
excitations with the modulation being generated by nonzero $\tilde{K}$ only as
it is evident from Eq.~\eqref{eq:dispBA} and the form of $\kappa_{\alpha\vc k}$. 
Flat dispersions are the fingerprints of underlying frustrations, and
resemble the Kitaev-Heisenberg model magnons characterized by two different
energy scales \cite{Cha13a}. The excitations in Eq.~\eqref{eq:dispBA} have two
branches for each triplon color $\alpha$ that cover the entire Brillouin zone
associated with the completed triangular lattice [dotted hexagon in
Fig.~\ref{fig:honeyKKt}(c)] by periodic copies of the smaller Brillouin zone
of the honeycomb lattice [full hexagon in Fig.~\ref{fig:honeyKKt}(c)].
The intensity of these excitations in the dynamic spin susceptibility varies
through the Brillouin zone -- while the upper branch dominates around the $\vc
q=\Gamma$ point, the lower branch is most intense around the AF wavevector
$\vc q=\Gamma'$. At the latter point (equivalent to $\vc q=\Gamma=0$ in terms
of the bosonic excitations), the magnetic excitations will eventually touch
zero energy signaling the transition into long-range AF phase as $\tilde K$
increases. This is also observed near the isotropic limit [see
Fig.~\ref{fig:honeyKKt}(e)] where the modulation of the originally flat
dispersions by the complementary interaction $\tilde{K}$ leads to a merging of
the two excitation branches and the result starts to resemble the excitonic
magnon dispersion. In contrast to the Heisenberg model at the same lattice, it
is characterized by a maximum at $\Gamma$ point, as has been seen
experimentally in the square-lattice case of \mbox{Ca$_2$RuO$_4$}
\cite{Jai17}.

The observed features of the presented gas-phase spectra close to the AF
transition are expected to be already quite indicative for the AF phase.
After the condensation, an additional excitation branch corresponding to the
amplitude (Higgs) mode will develop [a hint of this can be noticed in the
chain case when comparing Fig.~\ref{fig:chainKJ}(d) and (g)]. Besides that,
there will be also an ongoing redistribution of the spectral weight in the
magnon branch. 



\section{Conclusions}
\label{sec:concl}

We have studied singlet-triplet models that describe magnetism of spin-orbit
coupled $d^4$ Mott insulators, such as ruthenium Ru$^{4+}$ or iridium
Ir$^{5+}$ compounds. Singlet-triplet models appear in various physical
contexts (see, e.g., Refs.~\cite{Gia08,Som01,Che09,Cha13}) and are of general
interest because they host -- by very construction -- a quantum phase
transition from triplon gas to the ordered state of soft moments, when
exchange interactions overcome a singlet-triplet spin gap. The models
considered here bring a new feature into this physics: a magnetic frustration
that originates from bond-directional nature of orbital interactions
\cite{Kha05}. Similar to the case of spin-orbit pseudospin $J=1/2$ Mott
insulators with bond-directional Ising couplings
\cite{Jac09,Cha10,Liu18,San18} on a honeycomb lattice, the orbital frustration
has a strong impact on magnetism of singlet-triplet models \cite{Kha13,Ani19}.

The main aim of the present work was to understand how a triplon gas evolves
into a dense system of strongly interacting particles, in particular when
bond-directional anisotropy of the exchange interactions are taken to the
extreme as in Kitaev model. We find that this evolution is continuous and
results in a strongly correlated paramagnet smoothly connected to a triplon
gas. Even though this paramagnet misses the defining features of genuine
spin-liquids (many-body entanglement and emergent quasiparticles)
\cite{Sav17}, it is far from being trivial. In contrast to a conventional O(3)
singlet-triplet systems, spin correlations here are highly anisotropic and
strictly short-ranged even in the limit of strong exchange interactions where
the spin gap is very small. As in the Kitaev model, these peculiar features of
spin correlations follow from the symmetry properties of the model -- an
extensive number of conserved $Z_2$ quantities that decompose the Hilbert
space into subspaces with fixed bond-parity configurations. We have also shown
that the model can be mapped to a bilayer version of Kitaev model, but with
some additional terms in the interlayer couplings which act to suppress
gas-to-liquid phase transition in a bilayer Kitaev model
\cite{Tom18,Sei18,Tom19}. Exact diagonalization of the model in 1D-zigzag
chain as well as on honeycomb lattice show that the lowest energy excitations
are in the spin sector (and always gapped). This is different from the Kitaev
model with Majorana bands within the spin gap. 

Going away from Kitaev-like symmetry of the exchange interactions towards
isotropic O(3) limit, we find that triplons condense into AF ordered phase at
finite critical value of non-Kitaev $\tilde K$ term. At $\tilde K<K$ regime
(and close to phase transition), spin excitations show two distinct branches
of weakly dispersive modes [Fig.~\ref{fig:honeyKKt}(d)]; however, this result
is obtained within a mean-field treatment of the hard-core constraint
neglecting any multi-particle scattering processes. A quantitative description
of the highly frustrated magnetic condensate and its excitations in the regime
of $\tilde K<K$ remains an open theoretical problem.

Considering the model at negative $E_T$ values, we observe the links to some
exotic models such as biquadratic spin-1 and quantum dimer models. At negative
$E_T$ and small $\tilde K$ regime, we find a quantum phase transition from
strongly correlated paramagnetic phase to a plaquette-VBS state of the
triplon-dimers. 

Apart from a theoretical interest in frustrated singlet-triplet models, this
study was partly motivated by magnetic properties of honeycomb lattice
ruthenium compounds, in particular \mbox{Ag$_3$LiRu$_2$O$_6$}
\cite{Kim10,Kum19,Tak20}.  This compound is derived from \mbox{Li$_2$RuO$_3$} by
substituting Ag-ions for those Li-ions which reside between the Ru-honeycomb
planes. As a result, a structural transition observed in \mbox{Li$_2$RuO$_3$}
\cite{Miu07} is completely suppressed, suggesting \mbox{Ag$_3$LiRu$_2$O$_6$}
as a nearly ideal honeycomb lattice system to study the interplay between
spin-orbit coupling and exchange interactions. Current data \cite{Kim10,Kum19,Tak20}
shows that this compound is insulating and has no magnetic order, which would
be consistent with the (correlated) triplon-gas phase where interactions are
either too weak to overcome the spin-orbit gap, or they are strongly
frustrated preventing triplon condensation. As $4d$-electron wavefunctions are
rather extended in space, a direct overlap processes can be sizable in
ruthenates \cite{Str17}, thus raising the possibility of bond-directional
triplon dynamics in this material. Future experiments probing magnetic
dynamics are necessary to identify symmetry of the dominant exchange
interactions in \mbox{Ag$_3$LiRu$_2$O$_6$}. Metallic states induced by
electron doping of this material could bring some surprises as well. 

Overall, the orbitally frustrated singlet-triplet models show a rich physics,
interesting theoretically and also relevant to spin-orbit coupled
$J_\mathrm{eff}=0$ Mott insulators based on, e.g., ruthenium Ru$^{4+}$ and
iridium Ir$^{5+}$ ions. 


\acknowledgments

We would like to thank P.~Anisimov, M.~Daghofer, and T.~Takayama for useful
discussions.
J.Ch. acknowledges support by Czech Science Foundation (GA\v{C}R) under
Project No.~GA19-16937S and M\v{S}MT \v{C}R under NPU II project CEITEC 2020
(LQ1601). Computational resources were supplied by the Ministry of Education,
Youth and Sports of the Czech Republic under the Projects CESNET (Project No.
LM2015042) and CERIT-Scientific Cloud (Project No. LM2015085) provided within
the program Projects of Large Research, Development and Innovations
Infrastructures.
G.Kh. acknowledges support by the European Research Council under Advanced
Grant 669550 (Com4Com).




\end{document}